\documentclass[preprint,12pt,a4paper]{elsarticle}

\makeatletter
\def\ps@pprintTitle{%
  \let\@oddhead\@empty
  \let\@evenhead\@empty
  \def\@oddfoot{\reset@font\hfil\thepage\hfil}
  \let\@evenfoot\@oddfoot
}
\makeatother

\usepackage{amssymb}

\usepackage{hyperref}
\usepackage{breakurl}

\usepackage{amsmath}

\begin{document}

\begin{frontmatter}




\title{Scaling of global properties of fluctuating and mean streamwise velocities in pipe flow: Characterisation of a high Reynolds number transition region}


\author[au1]{Nils T. Basse}
\ead{nils.basse@npb.dk}

\address[au1]{Trubadurens v\"ag 8, 423 41 Torslanda, Sweden \\ \vspace{10 mm} \small {\rm \today}}


\begin{abstract}
We study the global, i.e. radially averaged, high Reynolds number (asymptotic) scaling of streamwise turbulence intensity squared defined as ${I^2=\overline{u^2}/U^2}$, where $u$ and $U$ are the fluctuating and mean velocities, respectively (overbar is time averaging). The investigation is based on the mathematical abstraction that the logarithmic region in wall turbulence extends across the entire inner and outer layers. Results are matched to spatially integrated Princeton Superpipe measurements [Hultmark M, Vallikivi M, Bailey SCC and Smits AJ. Logarithmic scaling of turbulence in smooth- and rough-wall pipe flow. J. Fluid Mech. {\bf 728}, 376-395 (2013)]. Scaling expressions are derived both for log-law and power-law functions of radius. A transition to asymptotic scaling is found at a friction Reynolds number $Re_{\tau} \sim 11000$.
\end{abstract}

\end{frontmatter}



\section{Introduction}

We have studied the scaling of streamwise turbulence intensity (TI) with Reynolds number in \cite{russo_a,basse_b,basse_c} and continue our research in this paper. We define the square of the TI as $I^2=\overline{u^2}/U^2$, where $u$ and $U$ are the fluctuating and mean velocities, respectively (overbar is time averaging). Physically, the square of the TI is equal to the ratio of the fluctuating and mean kinetic energy. The square of the normalised fluctuating velocity is $\overline{u^2}/U_{\tau}^2$, where $U_{\tau}$ is the friction velocity. In the literature, the square of the normalised fluctuating velocity is sometimes called the TI. The square of the normalised mean velocity is $U^2/U_{\tau}^2$.

As before, we use publicly available measurements from the Princeton Superpipe \cite{hultmark_a,smits_a}.

In previous work, we have studied the global, i.e. radially averaged TI and attempted to fit the measurements to equations with two parameters, either log- or power-laws. We have assumed that the same scaling expression is valid for all Reynolds numbers measured.

Now, we go one step further and assume that $\overline{u^2}/U_{\tau}^2$ and $U^2/U_{\tau}^2$ can be expressed separately as two-parameter functions of the distance from the wall, either as log-laws or power-laws. This approach provides the parameters as a function of Reynolds number, thereby allowing us to answer the question whether they depend on Reynolds number or not. The log-law is based on the description in \cite{marusic_a} and the power-law has a functional form which is equivalent to the log-law in the sense that:

\begin{equation}
x^{1/a} = \exp (\log(x)/a) \simeq 1 + \frac{\log(x)}{a},
\end{equation}

\noindent where $a$ is a constant, $x$ is a variable and the approximation is valid for $\log(x)/a \ll 1$. We make the assumption that the log-law and power-law hold across the entire inner and outer layers. For the log-law, this assumption is known to be unphysical outside the logarithmic region.

An important motivation for our work remains applications to computational fluid dynamics (CFD) simulations \cite{versteeg_a}, where the TI is often used as a boundary condition. Recently, a machine learning deep neural network has been applied to estimate the TI from short duration velocity signals \cite{corbetta_a}. This approach may lead to novel information on the temporal variation of the TI scaling behaviour.

The paper is organized as follows: In Section \ref{sec:TI_CFD}, we briefly summarize the importance of the TI as a boundary condition for CFD simulations. Thereafter, we review the local logarithmic velocity scaling in Section \ref{sec:local}, followed by global log-law and power-law scaling results in Section \ref{sec:global}. We discuss our findings in Section \ref{sec:discussion} and conclude in Section \ref{sec:conclusions}.

\section{Turbulence intensity as a CFD boundary condition}
\label{sec:TI_CFD}

A canonical example of a turbulence model where the TI is used as a boundary condition is the standard $k-\varepsilon$ model \cite{launder_a} where two equations are solved, one for the turbulent kinetic energy (TKE) $k$ and another for the rate of dissipation of the TKE, $\varepsilon$:

\begin{equation}
k=\frac{3}{2} (U_{\rm ref} I)^2
\end{equation}

\begin{equation}
\varepsilon=C_{\mu}^{3/4} \frac{k^{3/2}}{l},
\end{equation}

\noindent where $U_{\rm ref}$ is a characteristic velocity, $I^2=\overline{u^2}/U_{\rm ref}^2$ is the TI, $C_{\mu}$ is a dimensionless constant and $l$ is a characteristic length. We return to length scales in Section \ref{subsec:length_scales}.

Thus, we see that a reliable estimate of the TI is needed to calculate the TKE.

The kinematic turbulent viscosity is assumed to be isotropic and defined as:

\begin{equation}
\nu_t = C_{\mu} \frac{k^2}{\varepsilon} = C_{\mu}^{1/4} l \sqrt{3/2} U_{\rm ref} I
\end{equation}

An expression often used for the TI is one contained in the documentation of a commercial CFD software \cite{ansys_a}:

\begin{equation}
\label{eq:TI_ansys}
I=0.16 \times Re_{D_H}^{-1/8},
\end{equation}

\noindent where $Re_{D_H}$ is the Reynolds number based on the hydraulic diameter $D_H$. No reference is supplied, but the statement above the equation reads:

"The turbulence intensity at the core of a fully-developed duct flow can be estimated from the following formula derived from an empirical correlation for pipe flows:"

Note that other commercial (and open-source) CFD codes also use Equation (\ref{eq:TI_ansys}), e.g. \cite{siemens_a}. We find that problematic, since the origins of the equation is unclear and no references are provided. Our research aims to remedy the situation by modelling the TI using publicly available measurements.

\section{Local scaling}
\label{sec:local}

\begin{figure}[!ht]
\centering
\includegraphics[width=12cm]{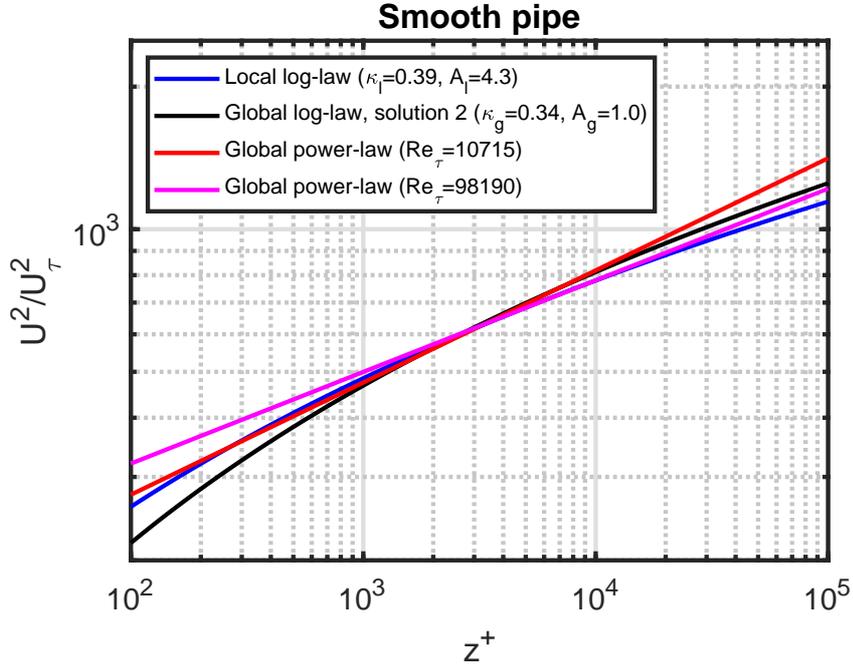}
\caption{Square of the normalised mean velocity as a function of $z^+$. The blue line shows Equation (\ref{eq:u_plus_sq}). The other lines are defined and referred to later in this paper: Sections \ref{sec:mean_log} and \ref{sec:mean_pow}.}
\label{fig:mean_sq_vs_z_plus}
\end{figure}

We first write the log-law for the streamwise mean velocity as formulated in \cite{marusic_a}:

\begin{eqnarray}
\label{eq:log_law}
  U^+_l(z) &=& \frac{1}{\kappa_l} \log (z^+) + A_l \\
   &=& \frac{1}{\kappa_l} \log(z/\delta) + \frac{1}{\kappa_l} \log(Re_{\tau}) + A_l,
\end{eqnarray}

\noindent where $U^+_l=U_l/U_{\tau}$, $U_l$ is the mean velocity in the streamwise direction, $z^+=z U_{\tau}/\nu$ is the normalized distance from the wall, $z$ is the distance from the wall, $\nu$ is the kinematic viscosity, $\kappa_l$ is the von K\'arm\'an constant and $A_l$ is a constant for a given wall roughness. Note that:

\begin{equation}
z/\delta = \frac{z^+}{Re_{\tau}},
\end{equation}

\noindent where $Re_{\tau}=\delta U_{\tau}/\nu$ is the friction Reynolds number and $\delta$ is the boundary layer thickness (pipe radius $R$ for pipe flow). The subscript "$l$" means that the constants are "local" fits, i.e. the range of $z$ where the log-law describes the measurements well (logarithmic region). Although the log-law is not valid close to the wall, we observe that $U^+_l=0$ if $z^+_l=\exp (-A_l \kappa_l)$. For the Princeton Superpipe constants ($A_l=4.3$ and $\kappa_l=0.39$), $z^+_l=0.18$, see Figure \ref{fig:mean_sq_vs_z_plus}.

In the following we will use the square of the log-law:

\begin{equation}
\label{eq:u_plus_sq}
\frac{U^2_l(z)}{U_{\tau}^2} = \frac{1}{\kappa_l^2} \log^2 (z^+) + A^2_l + \frac{2A_l}{\kappa_l} \log (z^+)
\end{equation}

\begin{figure}[!ht]
\centering
\includegraphics[width=12cm]{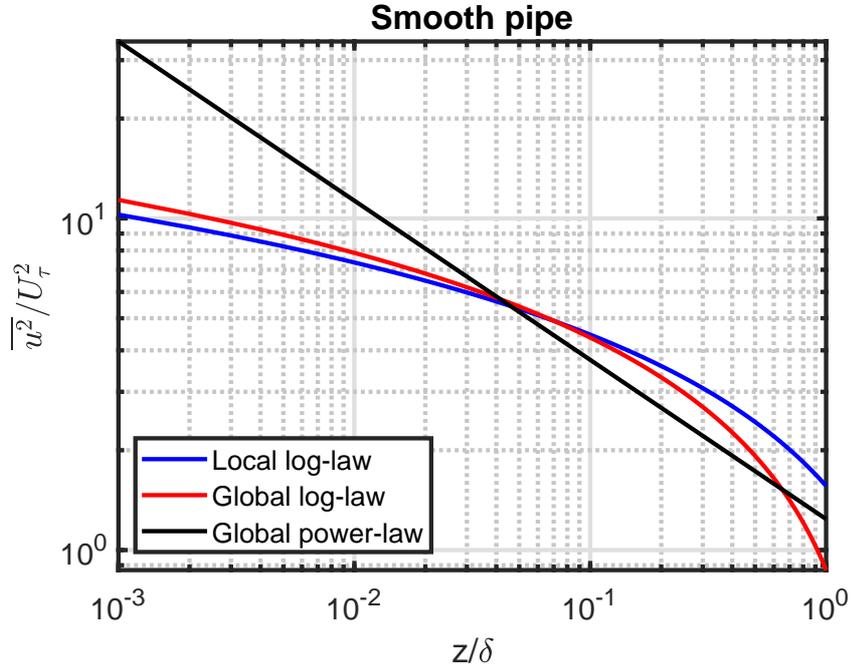}
\caption{Square of the normalised fluctuating velocity as a function of $z/\delta$. The blue line shows Equation (\ref{eq:u_rms_sq}). The other lines are defined and referred to later in this paper: Sections \ref{sec:fluc_log} and \ref{sec:fluc_pow}.}
\label{fig:turb_sq_vs_z_div_delta}
\end{figure}

According to the attached-eddy hypothesis prediction by Townsend \cite{townsend_a,marusic_a}, the streamwise fluctuating velocity $u_l$ can be written as:

\begin{equation}
\label{eq:u_rms_sq}
\frac{{\overline{u^2_l}}(z)}{U_{\tau}^2} = B_{1,l} - A_{1,l} \log (z/\delta),
\end{equation}

\noindent where $B_{1,l}$ and $A_{1,l}$ are constants. For the Princeton Superpipe, we use $B_{1,l}=1.56$ and $A_{1,l}=1.26$, see Figure \ref{fig:turb_sq_vs_z_div_delta}. We use $A_{1,l}=1.26$, which is an average of several datasets, including Superpipe measurements. The fit to Superpipe measurements gave $A_{1,l}=1.23 \pm 0.05$ \cite{marusic_a}.

We use Equations (\ref{eq:u_plus_sq}) and (\ref{eq:u_rms_sq}) to define the square of the turbulence intensity (TI):

\begin{equation}
\label{eq:TI_sq_def}
I^2_l(z) \rvert_{\rm log-law} =\frac{{\overline{u^2_l}}(z)}{U^2_l(z)}=\frac{B_{1,l} - A_{1,l} \log (z/\delta)}{\frac{1}{\kappa_l^2} \log^2 (z^+) + A^2_l + \frac{2A_l}{\kappa_l} \log (z^+)}
\end{equation}

Here, the two terms in the numerator are both positive and equal if $z/\delta$ is:

\begin{equation}
z/\delta \rvert_{0,l} = \exp (-B_{1,l}/A_{1,l})
\end{equation}

\begin{figure}[!ht]
\centering
\includegraphics[width=6.5cm]{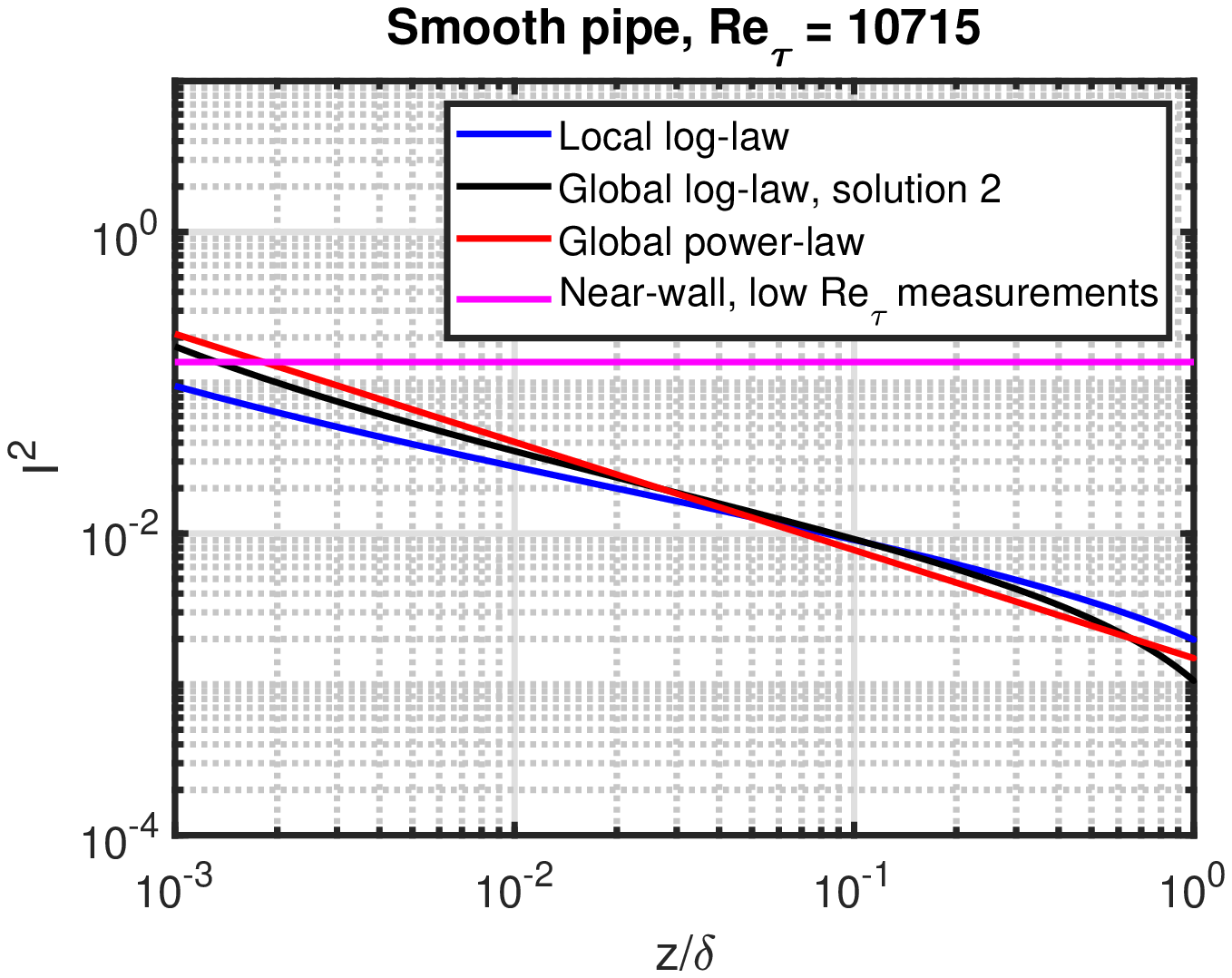}
\includegraphics[width=6.5cm]{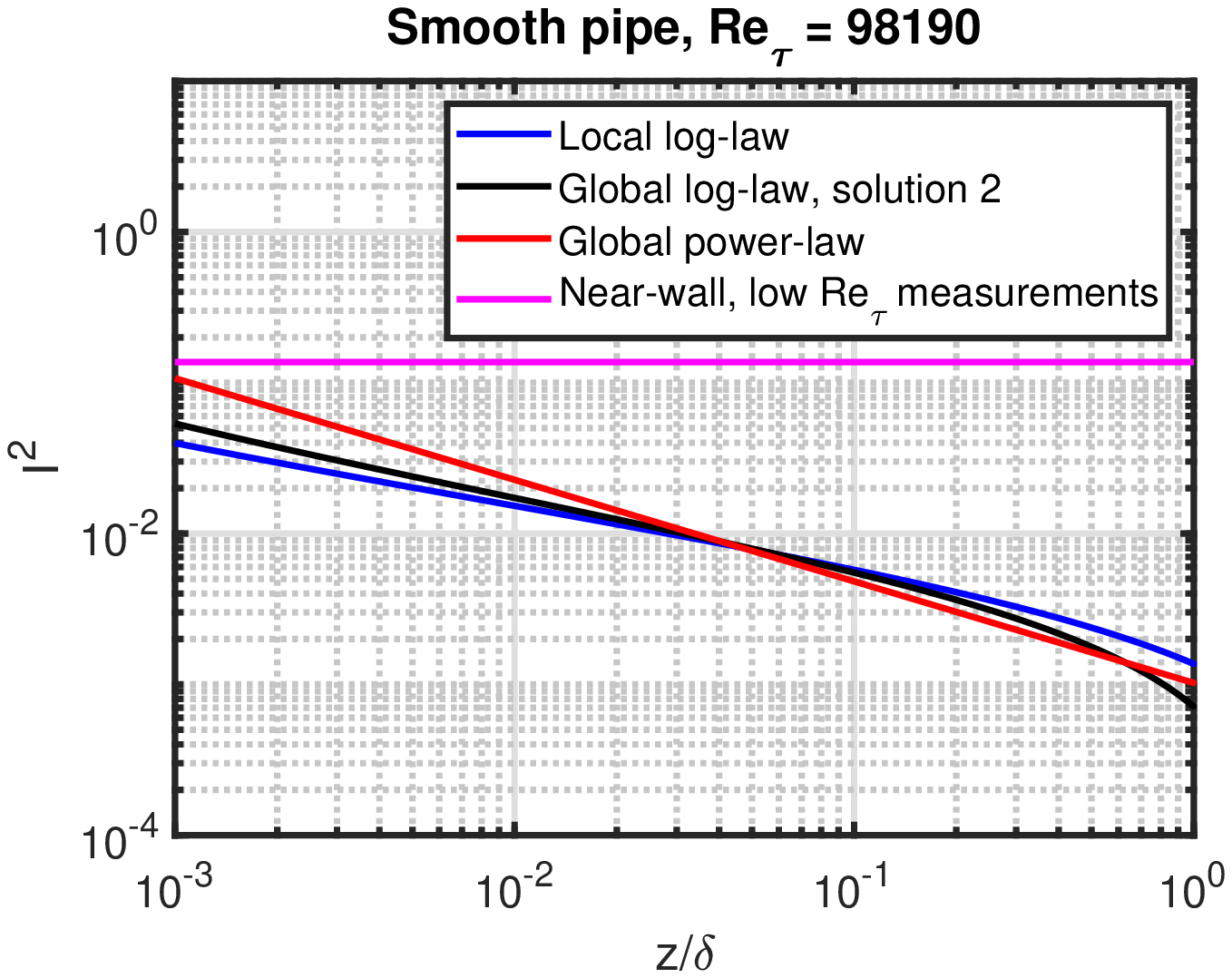}
\caption{$I^2$ as a function of $z/\delta$. Left: $Re_{\tau}=10715$, right: $Re_{\tau}=98190$. The blue lines show Equation (\ref{eq:TI_sq_def}). The magenta lines show the value $(0.37)^2$ which has been measured in the vicinity of the wall for low $Re_{\tau}$ turbulent pipe flow \cite{durst_a}. The other lines are defined and referred to later in this paper: Section \ref{sec:rad_prof}.}
\label{fig:I_sq_vs_alpha_lo_hi_Re}
\end{figure}

For the Superpipe constants, $z/\delta \rvert_{0,l} = 0.29$, see Figure \ref{fig:I_sq_vs_alpha_lo_hi_Re}; the measured near-wall low $Re_{\tau}$ $I^2$ from \cite{durst_a} is shown as a horizontal line for reference. We observe that the TI decreases with increasing $Re_{\tau}$.

For the Princeton Superpipe measurements, $Re_{\tau}$ ranges from 1985 to 98190; the logarithm of these numbers is 7.6 and 11.5, respectively.

\section{Global scaling}
\label{sec:global}

\subsection{Radial averaging definitions}

In this paper we make use of two radial averaging definitions, arithmetic mean and area-averaged. The reason for employing two methods is to extract two averages from the same measurements. This in turn allows us to construct radial profiles of the squared normalised fluctuating and mean velocities with two parameters. The result is two equations with two unknowns, i.e. the two profile parameters.

\subsubsection{Arithmetic mean}

The arithmetic mean (AM) is defined as:

\begin{eqnarray}
  \langle \cdot \rangle_{\rm AM} &=& \frac{1}{\delta} \int_{0}^{\delta} [\cdot] {\rm d}z \\
   &=& \frac{1}{Re_{\tau}} \int_{0}^{Re_{\tau}} [\cdot] {\rm d}z^+,
\end{eqnarray}

\noindent where we define the average both integrating over $z$ and $z^+$.

\subsubsection{Area-averaged}

The area-average (AA) is defined as:

\begin{eqnarray}
  \langle \cdot \rangle_{\rm AA} &=& \frac{2}{\delta^2} \int_{0}^{\delta} [\cdot] \times (\delta-z) {\rm d}z \\
   &=& \frac{2}{Re_{\tau}} \int_{0}^{Re_{\tau}} [\cdot] {\rm d}z^+ - \frac{2}{Re_{\tau}^2} \int_{0}^{Re_{\tau}} [\cdot] \times z^+ {\rm d}z^+
\end{eqnarray}

\subsection{Velocity fluctuations}

\begin{figure}[!ht]
\centering
\includegraphics[width=12cm]{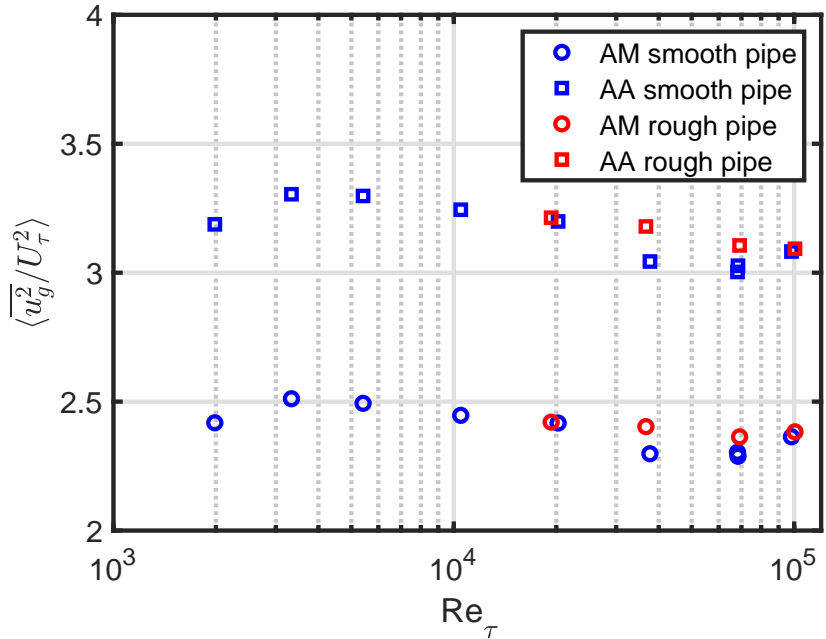}
\caption{The averaged square of the measured normalised fluctuating velocities as a function of Reynolds number.}
\label{fig:turb_sq_vs_Re_tau}
\end{figure}

The averaged square of the measured normalised fluctuating velocities is shown in Figure \ref{fig:turb_sq_vs_Re_tau}, both for smooth- and rough-wall pipe flow. Both the AM and AA averaging is shown; they have different amplitude, but both averages are roughly constant as a function of Reynolds number. We also note that the smooth- and rough-wall results are comparable.

In general, we show both smooth- and rough-wall measurements, but only use the smooth-wall measurements for postprocessing. The rough-wall measurements are shown for reference and to indicate whether they follow the smooth-wall behaviour or not.

\subsubsection{Log-law}
\label{sec:fluc_log}

The AM average of the log-law for the fluctuating velocity, Equation (\ref{eq:u_rms_sq}), has been derived in \cite{pullin_a}:

\begin{eqnarray}
  \biggl \langle \frac{{\overline{u^2_g}}}{U_{\tau}^2} \biggr \rangle_{\rm AM,log-law} &=& \frac{1}{\delta} \int_{0}^{\delta} [B_{1,g} - A_{1,g} \log (z/\delta)] {\rm d}z \\
   &=& B_{1,g} + A_{1,g} \label{eq:avrg_fluc_log_AM},
\end{eqnarray}

\noindent where the subscript "$g$" means that the parameters are "global", i.e. covering the entire range of $z$.

The corresponding AA average has been derived in \cite{basse_b}:

\begin{eqnarray}
  \biggl \langle \frac{{\overline{u^2_g}}}{U_{\tau}^2} \biggr \rangle_{\rm AA,log-law} &=& \frac{2}{\delta^2} \int_{0}^{\delta} [B_{1,g} - A_{1,g} \log (z/\delta)] \times (\delta-z) {\rm d}z \\
   &=& B_{1,g} + \frac{3}{2} \times A_{1,g} \label{eq:avrg_fluc_log_AA},
\end{eqnarray}

\noindent the difference being a factor $3/2$ multiplied with $A_{1,g}$.

\begin{figure}[!ht]
\centering
\includegraphics[width=6.5cm]{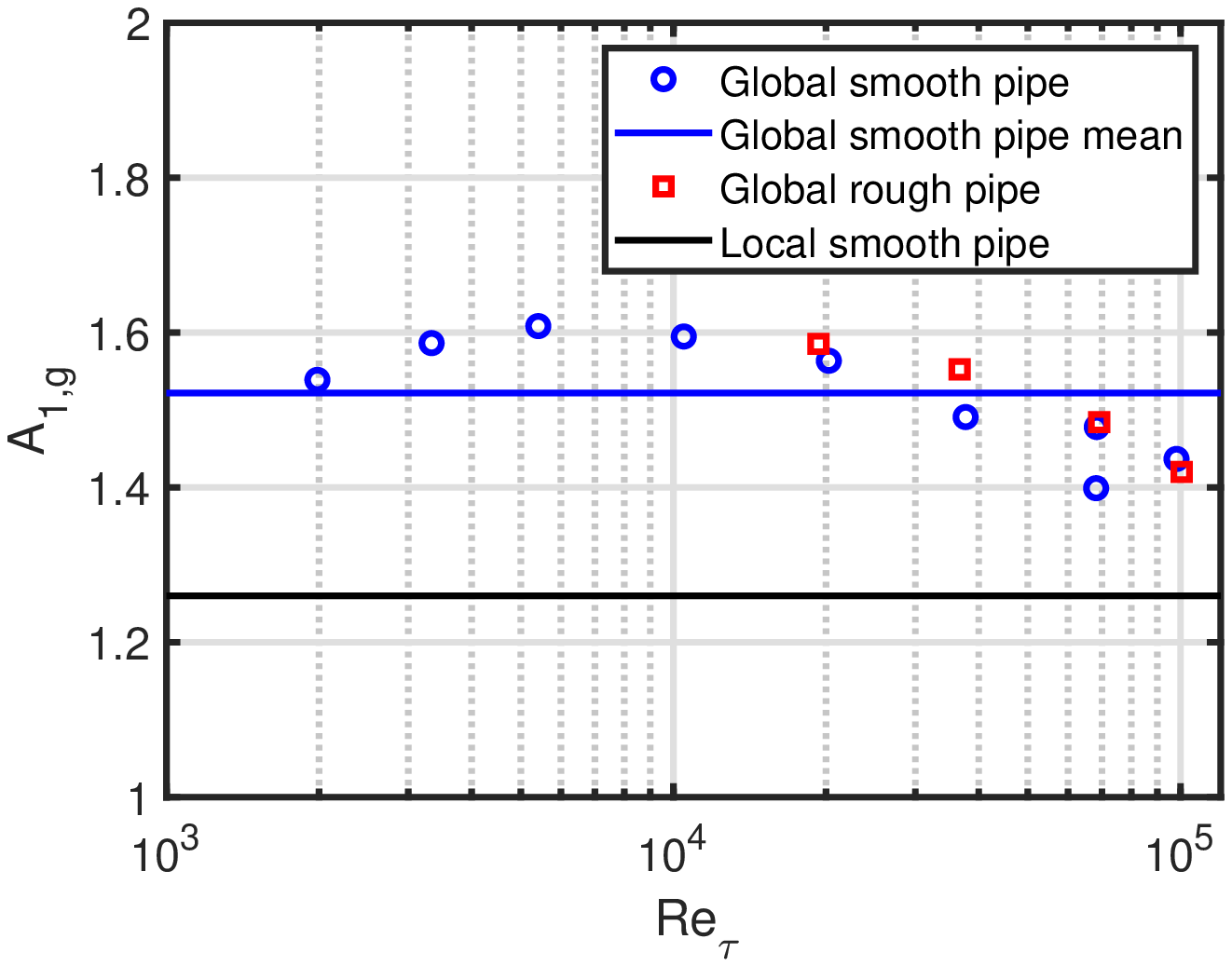}
\includegraphics[width=6.5cm]{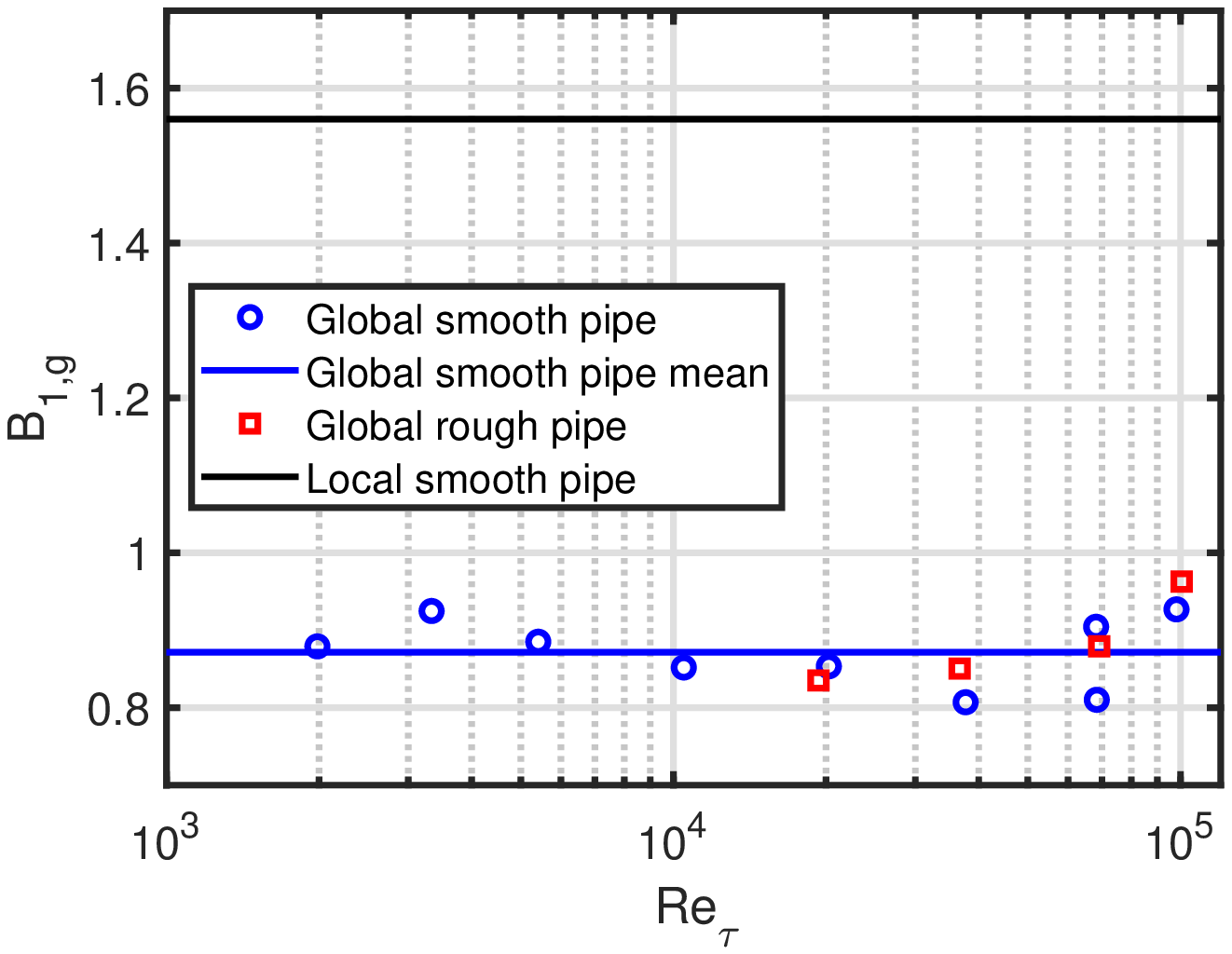}
\caption{Left-hand plot: $A_{1,g}$ vs. $Re_{\tau}$, right-hand plot: $B_{1,g}$ vs. $Re_{\tau}$. The smooth-wall average is shown as blue lines and the local parameter value as black lines.}
\label{fig:A_1_B_1_vs_Re}
\end{figure}

For each Reynolds number, the two averages can be used along with the measurements in Figure \ref{fig:turb_sq_vs_Re_tau} to derive $A_{1,g}$ and $B_{1,g}$, see Figure \ref{fig:A_1_B_1_vs_Re}. The mean and standard deviation is:

\begin{eqnarray}
  A_{1,g} &=& 1.52 \pm 0.07 \label{eq:cst_A_1g} \\
  B_{1,g} &=& 0.87 \pm 0.04 \label{eq:cst_B_1g},
\end{eqnarray}

\noindent compared to $A_{1,l}=1.26$ and $B_{1,l}=1.56$ for the local parameters. The global parameters can be used to calculate the AM and AA averages:

\begin{eqnarray}
  \biggl \langle \frac{{\overline{u^2_g}}}{U_{\tau}^2} \biggr \rangle_{\rm AM,log-law} &=& 2.39 \pm 0.08 \\
  \biggl \langle \frac{{\overline{u^2_g}}}{U_{\tau}^2} \biggr \rangle_{\rm AA,log-law} &=& 3.15 \pm 0.09, \label{eq:fluc_log}
\end{eqnarray}

\noindent where we have propagated the errors from Equations (\ref{eq:cst_A_1g}) and (\ref{eq:cst_B_1g}). The relative error for both the AM and AA averages is 3\%, which is comparable to the 3.4\% uncertainty of the Princeton Superpipe measurements, see Table 2 in \cite{hultmark_a}.

Results using the global parameters are shown in Figure \ref{fig:turb_sq_vs_z_div_delta} as the "Global log-law".

\subsubsection{Power-law}
\label{sec:fluc_pow}

In addition to the log-law for the fluctuating velocity, our alternative radial profile will be a power-law function which we write introducing two new parameters $a_g$ and $b_g$:

\begin{equation}
\frac{{\overline{u^2_g}}(z)}{U_{\tau}^2} = a_g \times \left( \frac{z}{\delta} \right)^{b_g}
\end{equation}

As we did for the log-law, we calculate the AM and AA averages of this function:

\begin{eqnarray}
  \biggl \langle \frac{{\overline{u^2_g}}}{U_{\tau}^2} \biggr \rangle_{\rm AM,power-law} &=& \frac{1}{\delta} \int_{0}^{\delta} \left[ a_g \times \left( \frac{z}{\delta} \right)^{b_g} \right] {\rm d}z \\
   &=&\frac{a_g}{b_g+1} \label{eq:avrg_fluc_pow_AM}
\end{eqnarray}

\begin{eqnarray}
  \biggl \langle \frac{{\overline{u^2_g}}}{U_{\tau}^2} \biggr \rangle_{\rm AA,power-law} &=& \frac{2}{\delta^2} \int_{0}^{\delta} \left[ a_g \times \left( \frac{z}{\delta} \right)^{b_g} \right] \times (\delta-z) {\rm d}z \\
   &=& \frac{2a_g}{(b_g+1)(b_g+2)} \label{eq:avrg_fluc_pow_AA}
\end{eqnarray}

\begin{figure}[!ht]
\centering
\includegraphics[width=6.5cm]{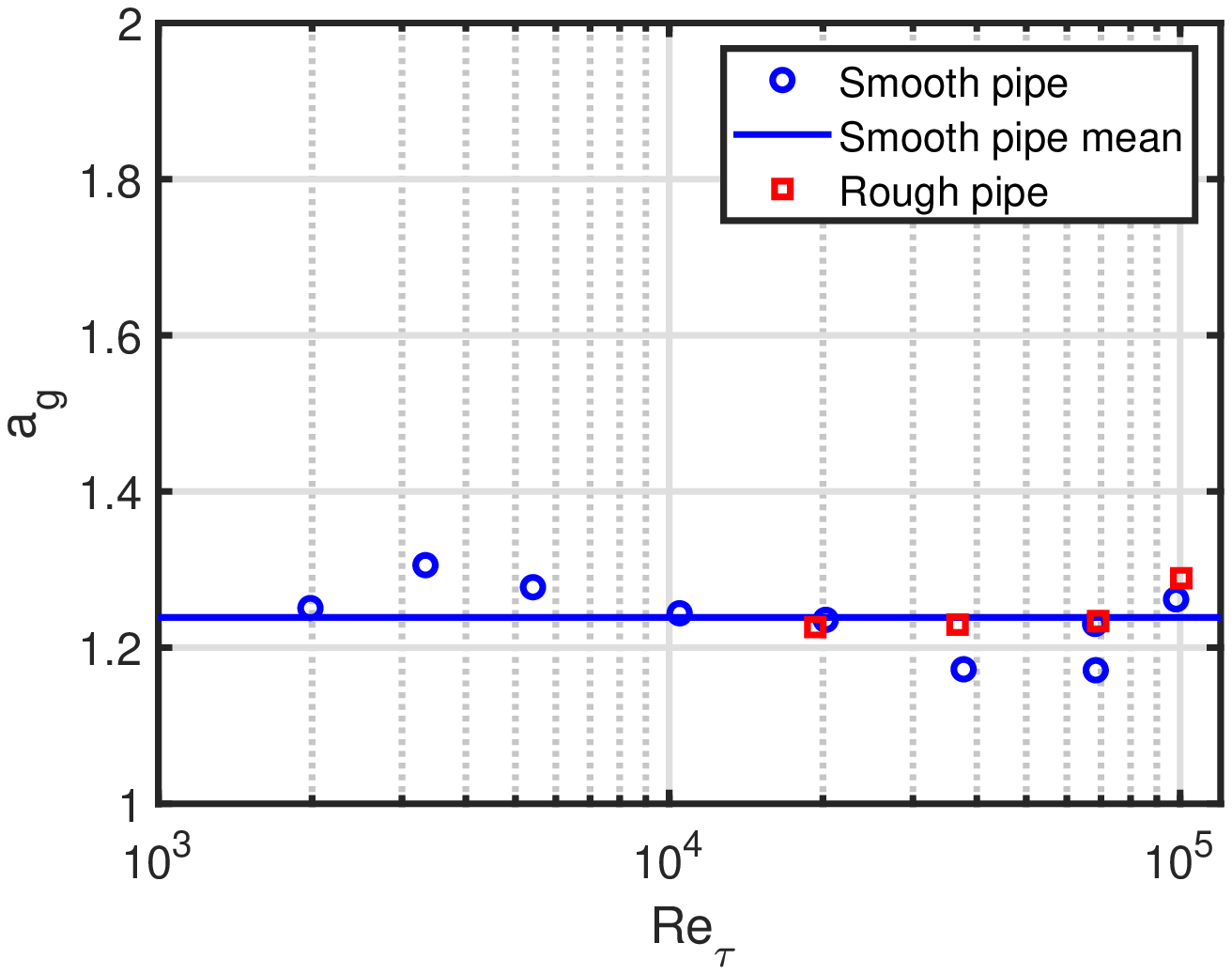}
\includegraphics[width=6.5cm]{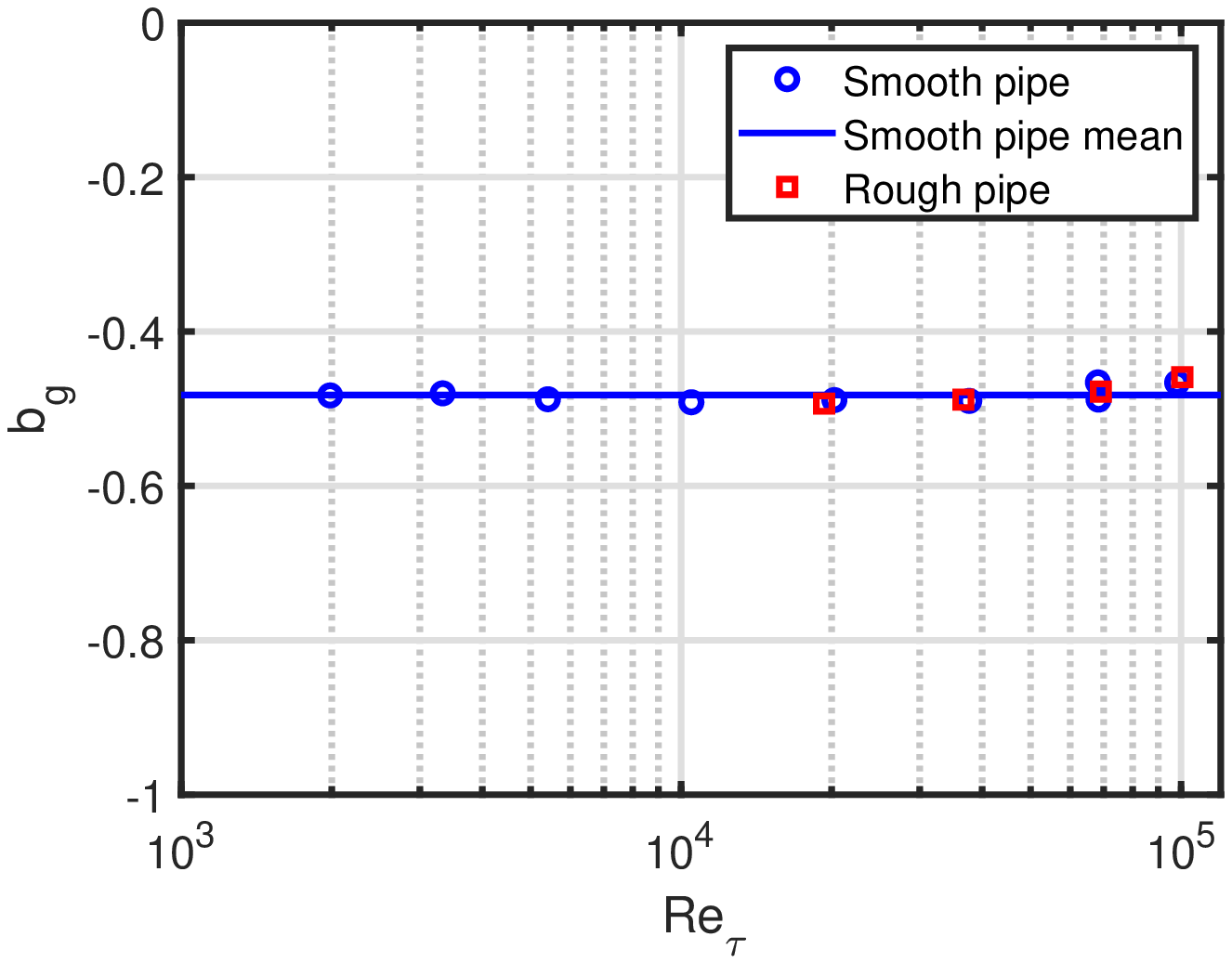}
\caption{Left-hand plot: $a_g$ vs. $Re_{\tau}$, right-hand plot: $b_g$ vs. $Re_{\tau}$. The smooth-wall average is shown as blue lines.}
\label{fig:small_a_b_vs_Re}
\end{figure}

As for the log-law, these two averaged equations can be used with the measurements in Figure \ref{fig:turb_sq_vs_Re_tau} to calculate $a_g$ and $b_g$ for each Reynolds number, see Figure \ref{fig:small_a_b_vs_Re}.

The mean and standard deviation using the smooth-wall measurements is:

\begin{eqnarray}
  a_g &=& 1.24 \pm 0.04 \\
  b_g &=& -0.48 \pm 0.01
\end{eqnarray}

By construction, the global power-law parameters yield (almost) the same result for the AM and AA averages as the global log-law parameters:

\begin{eqnarray}
  \biggl \langle \frac{{\overline{u^2_g}}}{U_{\tau}^2} \biggr \rangle_{\rm AM,power-law} &=& 2.38 \\
  \biggl \langle \frac{{\overline{u^2_g}}}{U_{\tau}^2} \biggr \rangle_{\rm AA,power-law} &=& 3.14 \label{eq:fluc_pow}
\end{eqnarray}

Results using the global power-law parameters are shown in Figure \ref{fig:turb_sq_vs_z_div_delta} as the "Global power-law". The shape of the power-law is very different from the log-law profiles, also in the range where the log-law matches measurements well. This is an indication that the power-law is far from reality; in that sense, it is a mathematical abstraction. However, the average of the profile does match the average of the measurements as is the case for the log-law.

\subsection{Mean velocity}

\begin{figure}[!ht]
\centering
\includegraphics[width=12cm]{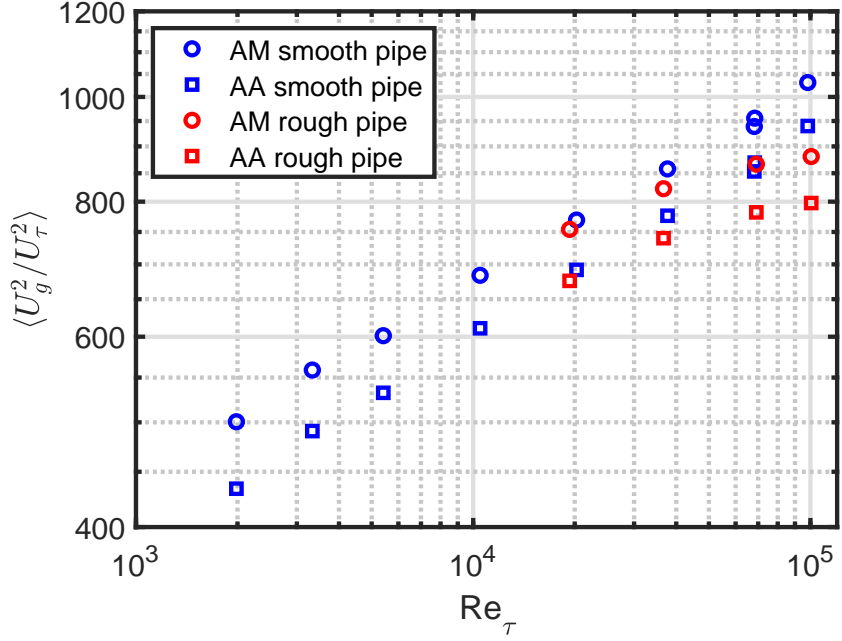}
\caption{The averaged square of the measured normalised mean velocities as a function of Reynolds number.}
\label{fig:mean_sq_vs_Re_tau}
\end{figure}

The averaged square of the measured normalised mean velocities is shown in Figure \ref{fig:mean_sq_vs_Re_tau}, both for smooth- and rough-wall pipe flow. Both the AM and AA averaging is shown; they increase with Reynolds number, but opposed to the fluctuating velocity, the amplitude of the AM averaging is higher than for the AA averaging. We can thus already conclude that the scaling of the TI is really due to the scaling of the mean velocity. For this case, the smooth- and rough-wall results deviate.

\subsubsection{Log-law}
\label{sec:mean_log}

As we did for the fluctuating velocities, we also average the square of the mean velocities. The AM average of the log-law has been derived in \cite{basse_a}:

\begin{eqnarray}
  \biggl \langle \frac{U^2_g}{U_{\tau}^2} \biggr \rangle_{\rm AM,log-law} &=& \frac{1}{Re_{\tau}} \int_{0}^{Re_{\tau}} \left[ \frac{1}{\kappa_g} \log(z^+)+A_g \right]^2 {\rm d}z^+ \\ \label{eq:avrg_mean_log_AM}
   &=& \frac{2}{\kappa_g^2} - \frac{2A_g}{\kappa_g} + A^2_g + \log(Re_{\tau}) \left( \frac{2A_g}{\kappa_g} - \frac{2}{\kappa^2_g} \right) \\ \nonumber
   & &  + \frac{\log^2(Re_{\tau})}{\kappa^2_g},
\end{eqnarray}

\noindent and the AA average is:

\begin{eqnarray}
  \biggl \langle \frac{U^2_g}{U_{\tau}^2} \biggr \rangle_{\rm AA,log-law} &=& \frac{2}{Re_{\tau}} \int_{0}^{Re_{\tau}} \left[ \frac{1}{\kappa_g}\log(z^+)+A_g \right]^2 {\rm d}z^+ \\ \nonumber
  & & - \frac{2}{Re_{\tau}^2} \int_{0}^{Re_{\tau}} \left[ \frac{1}{\kappa_g}\log(z^+)+A_g \right]^2 \times z^+ {\rm d}z^+ \\ \label{eq:avrg_mean_log_AA}
   &=& \frac{7}{2\kappa_g^2} - \frac{3A_g}{\kappa_g} + A^2_g + \log(Re_{\tau}) \left( \frac{2A_g}{\kappa_g} - \frac{3}{\kappa^2_g} \right) \\ \nonumber
   & &  + \frac{\log^2(Re_{\tau})}{\kappa^2_g}
\end{eqnarray}

\begin{figure}[!ht]
\centering
\includegraphics[width=6.5cm]{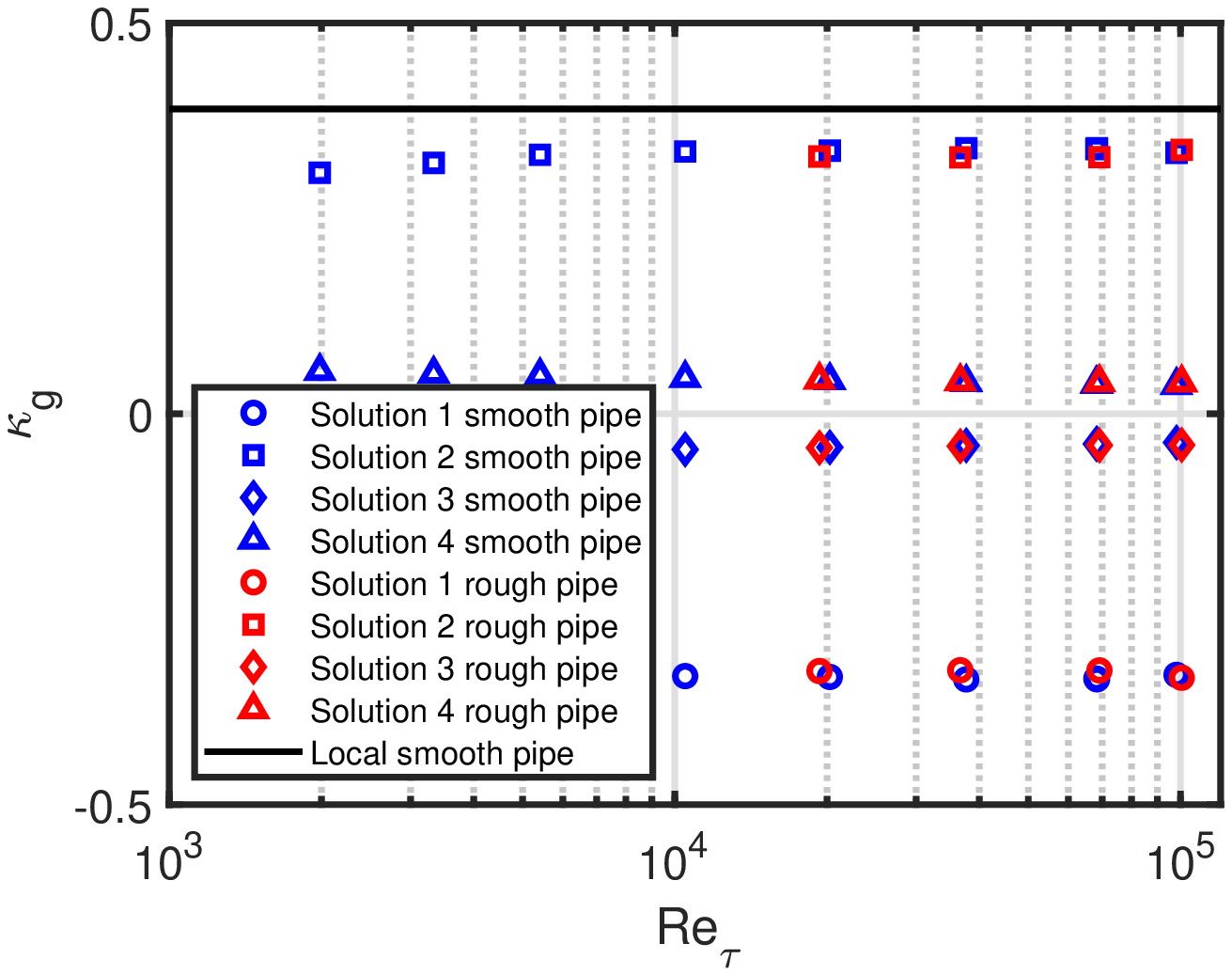}
\includegraphics[width=6.5cm]{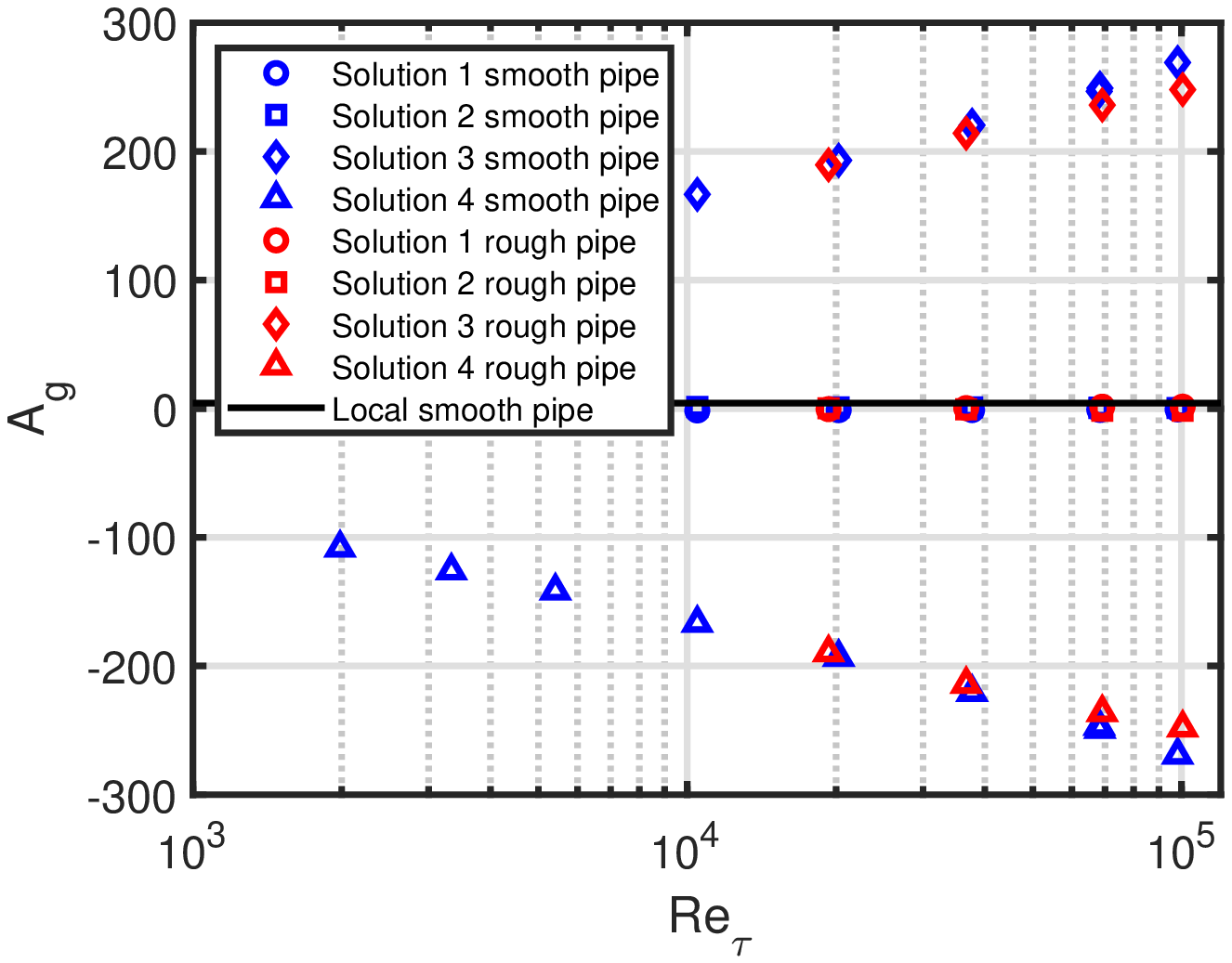}
\caption{Left-hand plot: $\kappa_g$ as a function of $Re_{\tau}$, right-hand plot: $A_g$ as a function of $Re_{\tau}$. The local parameter values are shown as black lines.}
\label{fig:kappa_A_vs_Re}
\end{figure}

\begin{figure}[!ht]
\centering
\includegraphics[width=6.5cm]{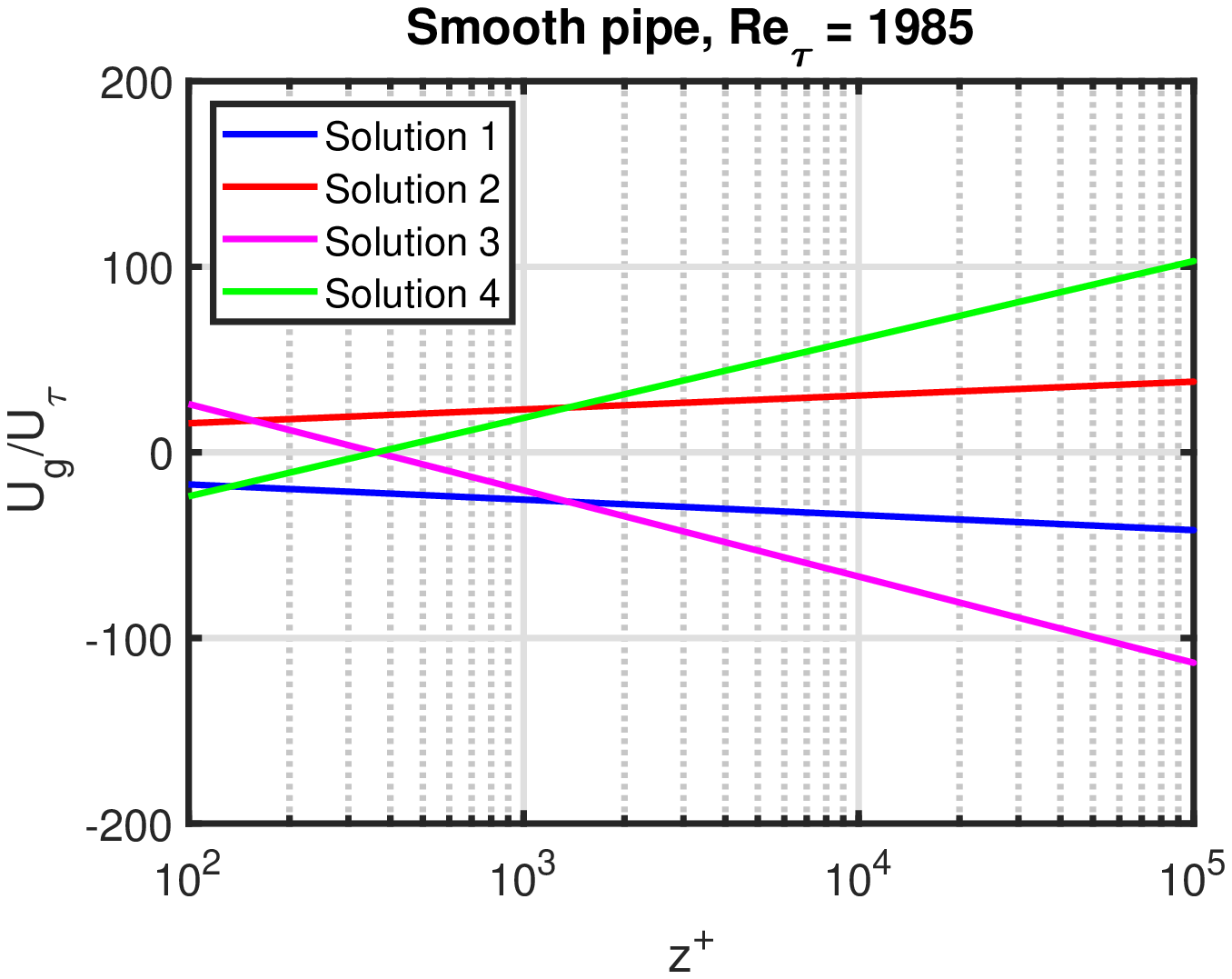}
\includegraphics[width=6.5cm]{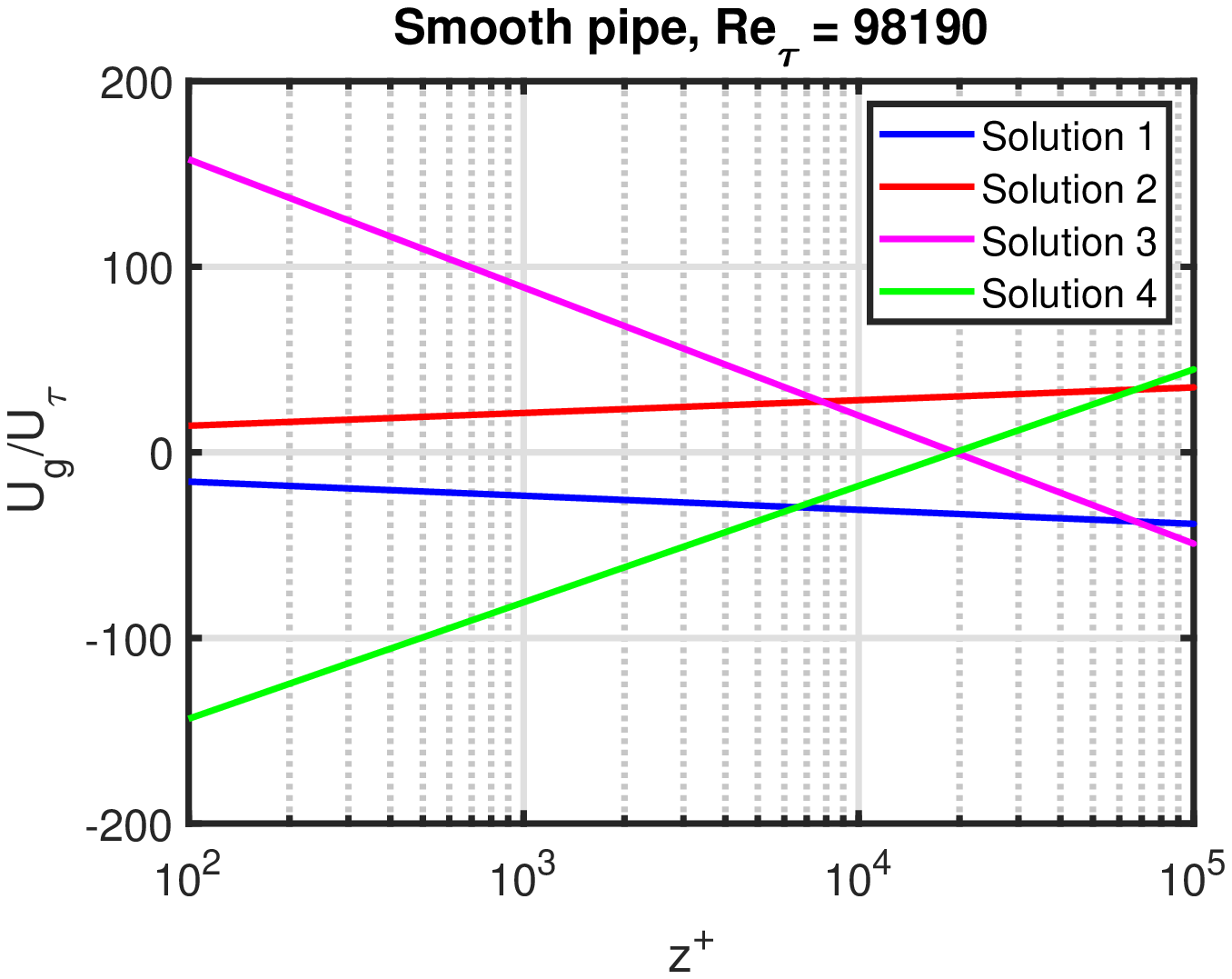}
\caption{Normalised mean velocity as function of $z^+$ for the four solutions. Left-hand plot: $Re_{\tau}=1985$, right-hand plot: $Re_{\tau}=98190$.}
\label{fig:solutions_lo_hi_Re}
\end{figure}

These two equations along with the averaged measurements form a system of two quadratic equations; thus, we have four solutions; $\kappa_g$ and $A_g$ for these solutions are shown in Figure \ref{fig:kappa_A_vs_Re}. Only two of the solutions are unique, the other two are mirror images, see Figure \ref{fig:solutions_lo_hi_Re}. Note that we show the normalised mean velocity; the square of this yields two unique solutions.

We focus on solution 2, which is the solution where $\kappa_g$ and $A_g$ are closest to $\kappa_l$ and $A_l$. $A_{g,{\rm solution~2}}$ does not vary with $Re_{\tau}$, whereas $\kappa_{g,{\rm solution~2}}$ does:

\begin{eqnarray}
  A_{g,{\rm solution~2}} &=& 1.01 \pm 0.32 \label{eq:sol2_kappa} \\
  \kappa_{g,{\rm solution~2}} &=& 0.34 - 623.9 \times Re_{\tau}^{-1.31} \qquad R^2=0.97 \label{eq:sol2_A}
\end{eqnarray}

\begin{figure}[!ht]
\centering
\includegraphics[width=6.5cm]{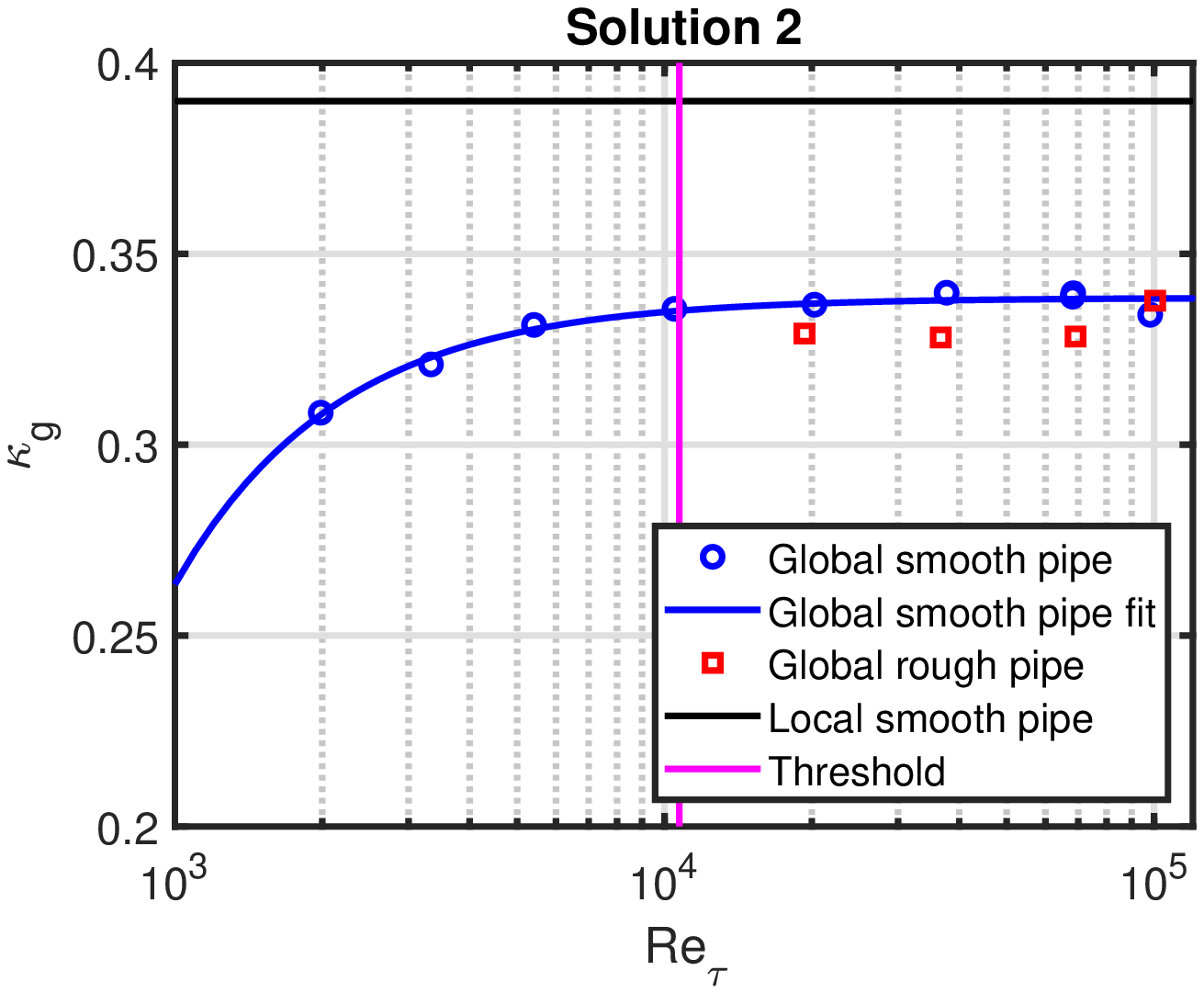}
\includegraphics[width=6.5cm]{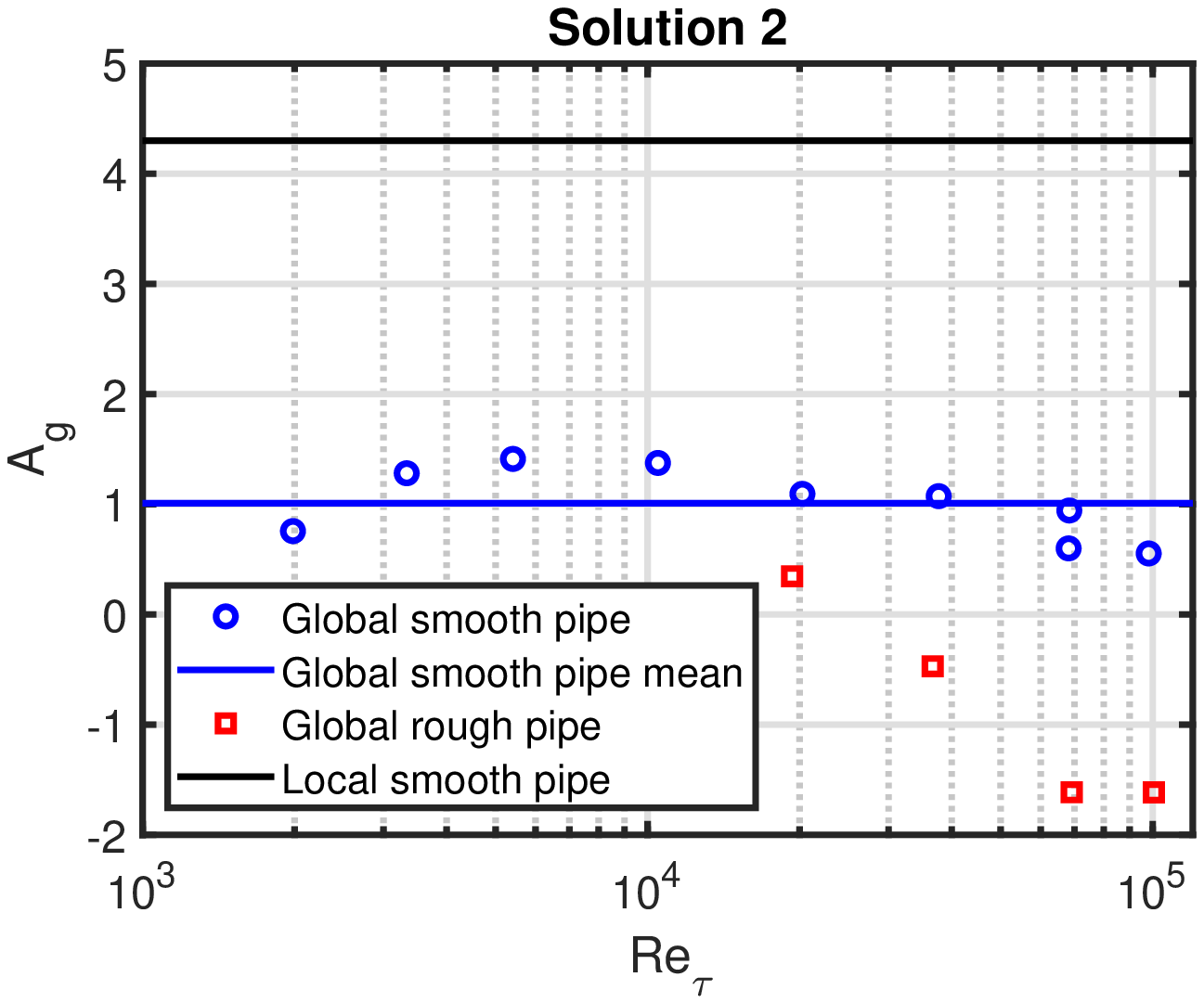}
\caption{Left-hand plot: $\kappa_{g,{\rm solution~2}}$ vs. $Re_{\tau}$, right-hand plot: $A_{g,{\rm solution~2}}$ vs. $Re_{\tau}$. Blue lines are Equations (\ref{eq:sol2_kappa}) and (\ref{eq:sol2_A}) and local parameter values are black lines.}
\label{fig:solution2_kappa_A}
\end{figure}

Solution 2 is provided as mean and standard deviation for $A_{g,{\rm solution~2}}$ and as a function of $Re_{\tau}$ for $\kappa_{g,{\rm solution~2}}$ along with the coefficient of determination $R^2$. Values are shown in Figure \ref{fig:solution2_kappa_A}. The asymptotic value for $\kappa_{g,{\rm solution~2}}$ is 0.34; this solution is shown as the "Global log-law, solution 2" in Figure \ref{fig:mean_sq_vs_z_plus}. The rough-wall pipe parameters deviate from the smooth-wall pipe results.

We define a $Re_{\tau}$ threshold from the $\kappa_{g,{\rm solution~2}}$ scaling which we determine as the value where 99\% of the asymptotic value of $\kappa_{g,{\rm solution~2}}$ is reached. This threshold value is $Re_{\tau} \rvert_{\rm threshold}=10715$, see the vertical magenta line in the left-hand plot of Figure \ref{fig:solution2_kappa_A}. This establishes a high Reynolds number transition region which will be a recurring topic in the remainder of this paper.

Solution 4, which has a small positive $\kappa_g$ and a large negative $A_g$ will be treated in the Discussion. It is a solution which has the zero crossing at much larger values of $z^+$ than solution 2 because of a large amplitude (but negative) $\kappa_g A_g$ product.

\begin{figure}[!ht]
\centering
\includegraphics[width=12cm]{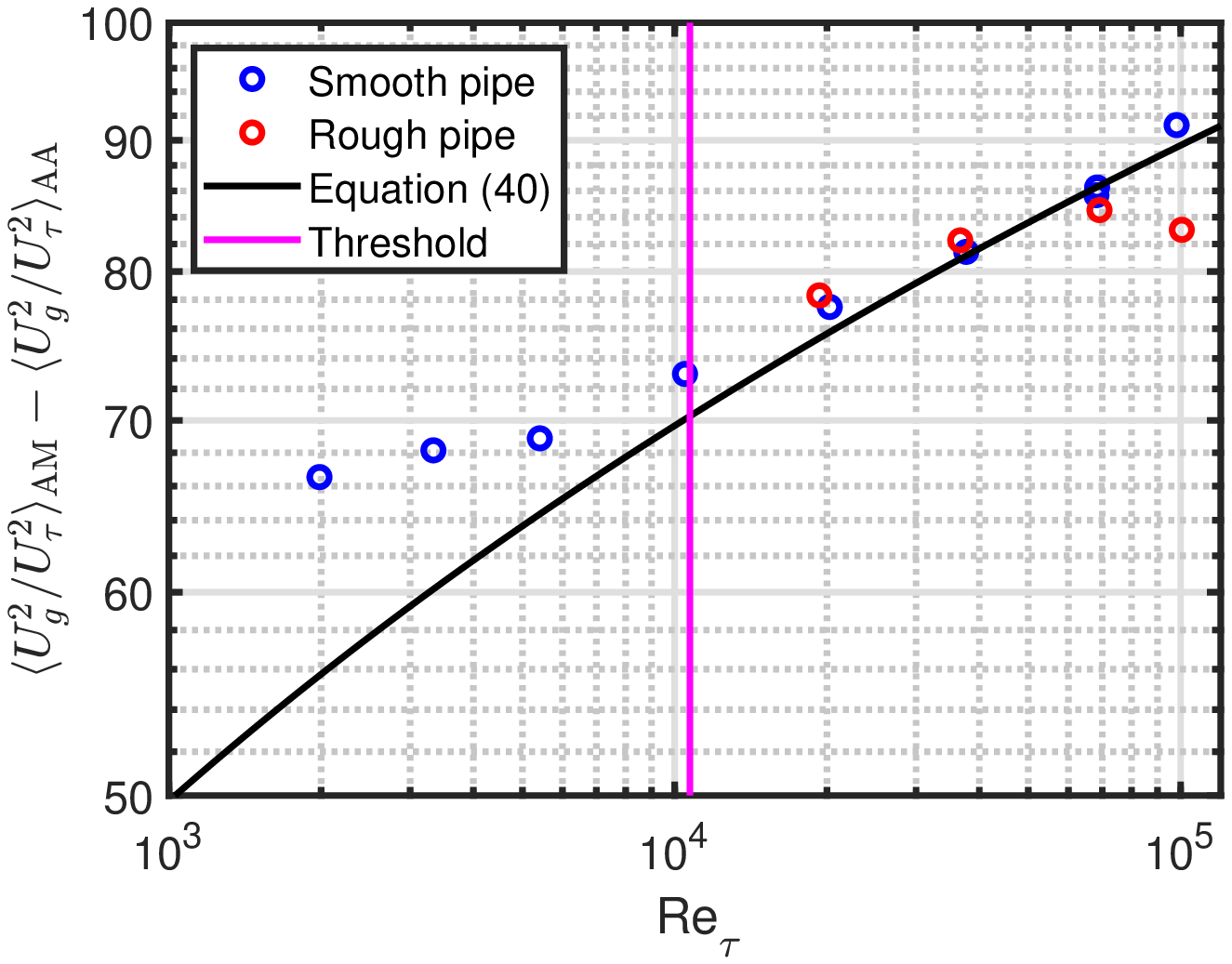}
\caption{$\biggl \langle \frac{U^2_g}{U_{\tau}^2} \biggr \rangle_{\rm AM}-\biggl \langle \frac{U^2_g}{U_{\tau}^2} \biggr \rangle_{\rm AA}$ vs. $Re_{\tau}$, the black line is Equation (\ref{eq:calc_mean_diff}).}
\label{fig:mean_am_aa_diff}
\end{figure}

The difference between Equations (\ref{eq:avrg_mean_log_AM}) and (\ref{eq:avrg_mean_log_AA}) is:

\begin{equation}
\label{eq:calc_mean_diff}
\biggl \langle \frac{U^2_g}{U_{\tau}^2} \biggr \rangle_{\rm AM,log-law}-\biggl \langle \frac{U^2_g}{U_{\tau}^2} \biggr \rangle_{\rm AA,log-law}=\frac{A_g}{\kappa_g} - \frac{3}{2 \kappa_g^2} + \log(Re_{\tau}) \left( \frac{1}{\kappa_g^2} \right),
\end{equation}

\noindent which is shown for $A_g=1.01$ (Equation (\ref{eq:sol2_kappa})) and $\kappa_g=0.34$ (Equation (\ref{eq:sol2_A})) in Figure \ref{fig:mean_am_aa_diff} along with the difference for the measurements. As expected, the difference scales with $\log(Re_{\tau})$ for Reynolds numbers above the threshold.

\subsubsection{Power-law}
\label{sec:mean_pow}

For the mean velocity, we also identify an alternative radial power-law profile with two new parameters $c_g$ and $d_g$ for the squared normalised mean velocity:

\begin{equation}
\label{eq:sq_mean_pow}
\frac{U_g^2(z)}{U_{\tau}^2} = c_g \times \left( z^+ \right)^{d_g}
\end{equation}

The AM and AA averages of Equation (\ref{eq:sq_mean_pow}) are:

\begin{eqnarray}
  \biggl \langle \frac{U^2_g}{U_{\tau}^2} \biggr \rangle_{\rm AM,power-law} &=& \frac{1}{Re_{\tau}} \int_{0}^{Re_{\tau}} \left[ c_g \times \left( z^+ \right)^{d_g} \right] {\rm d}z^+ \\
   &=& \frac{c_g}{d_g+1} \times Re_{\tau}^{d_g} \label{eq:avrg_mean_pow_AM}
\end{eqnarray}

\begin{eqnarray}
  \biggl \langle \frac{U^2_g}{U_{\tau}^2} \biggr \rangle_{\rm AA,power-law} &=& \frac{2}{Re_{\tau}} \int_{0}^{Re_{\tau}} \left[ c_g \times \left( z^+ \right)^{d_g}  \right] {\rm d}z^+ \\ \nonumber
  & & - \frac{2}{Re_{\tau}^2} \int_{0}^{Re_{\tau}} \left[ c_g \times \left( z^+ \right)^{d_g} \right] \times z^+ {\rm d}z^+ \\
   &=& \frac{2c_g}{(d_g+1)(d_g+2)} \times Re_{\tau}^{d_g} \label{eq:avrg_mean_pow_AA}
\end{eqnarray}

\begin{figure}[!ht]
\centering
\includegraphics[width=6.5cm]{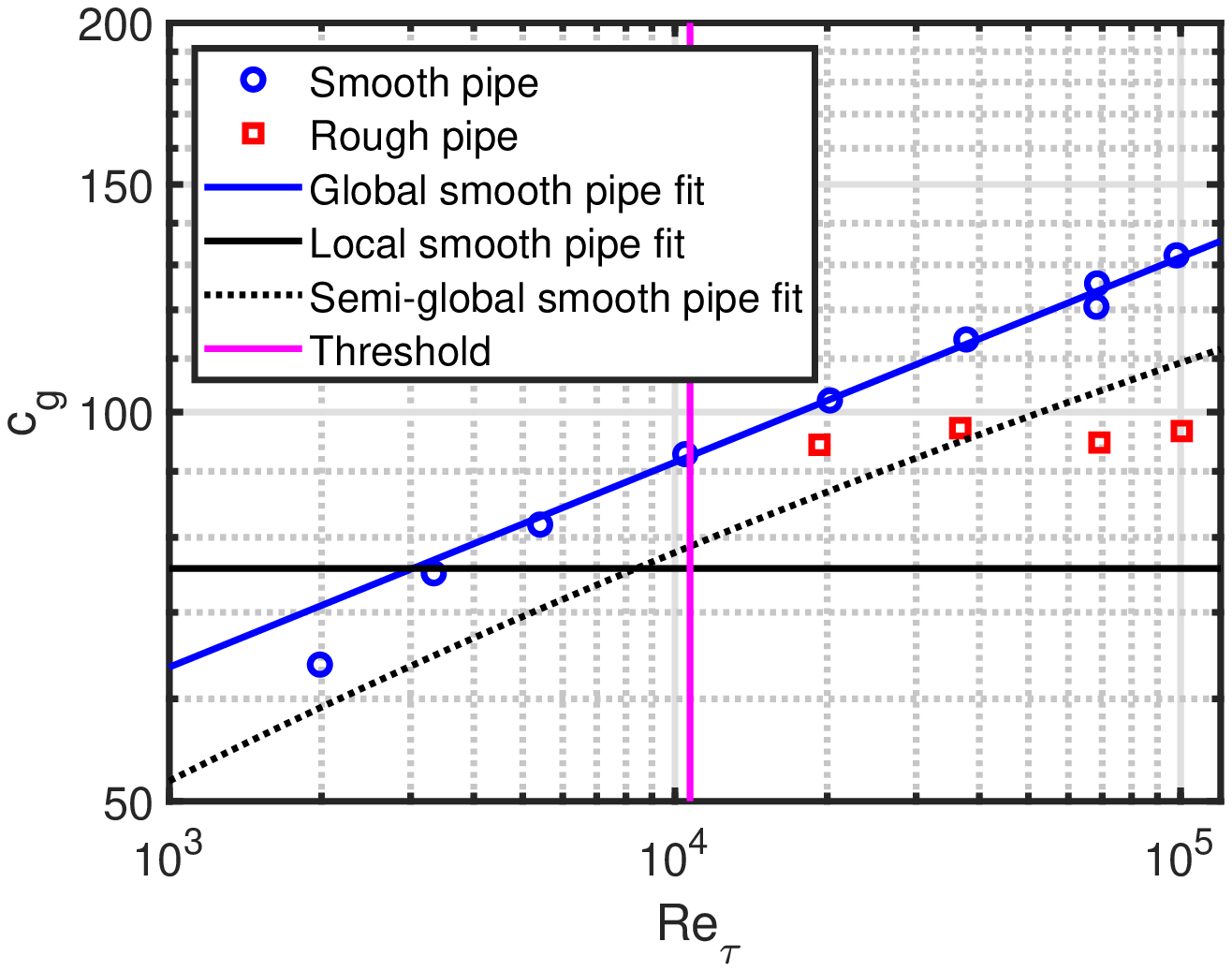}
\includegraphics[width=6.5cm]{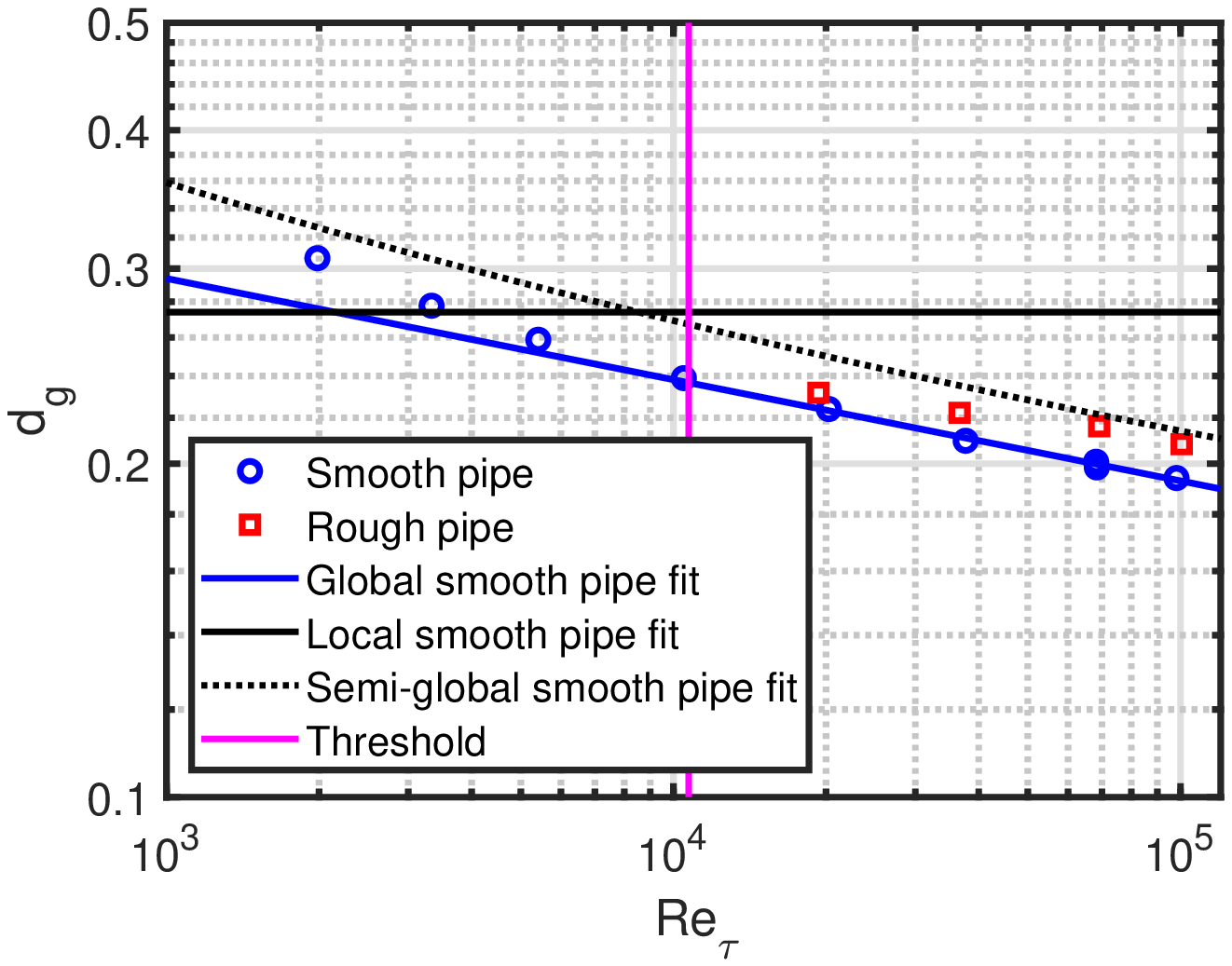}
\caption{Left-hand plot: $c_g$ as function of $Re_{\tau}$, right-hand plot: $d_g$ as function of $Re_{\tau}$. Blue lines are Equations (\ref{eq:c_g}) and (\ref{eq:d_g}), black solid lines are local values from Equation (\ref{eq:zag_l}) and black dotted lines are semi-global values from Equation (\ref{eq:zag_s-g}).}
\label{fig:small_c_d_vs_Re}
\end{figure}

From solving the equations it is clear that both $c_g$ and $d_g$ have a Reynolds number dependency, see Figure \ref{fig:small_c_d_vs_Re}. The rough-wall data differs from the smooth-wall solutions for the power-law profile as it did for the log-law.

We fit the parameters to power-laws of $Re_{\tau}$ above $Re_{\tau} \rvert_{\rm threshold}$:

\begin{eqnarray}
  c_g &=& c_1 \times Re_{\tau}^{c_2} \\
    &=& 21.25 \times Re_{\tau}^{0.16} \qquad R^2=0.97
\label{eq:c_g}
\end{eqnarray}

\begin{eqnarray}
  d_g &=& d_1 \times Re_{\tau}^{d_2} \\
    &=& 0.55 \times Re_{\tau}^{-0.09} \qquad R^2=0.99
\label{eq:d_g}
\end{eqnarray}

The solution is shown for $Re_{\tau}=10715$ and $Re_{\tau}=98190$ as the "Global power-law" in Figure \ref{fig:mean_sq_vs_z_plus}.

Also included in Figure \ref{fig:small_c_d_vs_Re} are previous results for local and semi-global (subscript "s-g") power-law fits to Superpipe measurements \cite{zagarola_a,zagarola_b}:

\begin{equation}
\label{eq:zag_l}
\frac{U_{\rm l}^2(z)}{U_{\tau}^2} = \left[ 8.70 \times (z^+)^{0.137} \right]^2
\end{equation}

\begin{equation}
\label{eq:zag_s-g}
\frac{U_{\rm s-g}^2(z)}{U_{\tau}^2} = \left[ (0.7053 \times \log (Re_D) + 0.3055) \times (z^+)^{\frac{1.085}{\log (Re_D)}+\frac{6.535}{\log^2(Re_D)}} \right]^2,
\end{equation}

\noindent where $Re_D=D \langle U_g \rangle_{\rm AA}/\nu$ is the bulk Reynolds number based on the pipe diameter $D=2R$. We use an equation derived in \cite{basse_c} to convert between $Re_{\tau}$ and $Re_D$:

\begin{equation}
\label{eq:re_d_to_re_tau}
Re_{\tau} = 0.0621 \times Re_D^{0.9148}
\end{equation}

Note that the exponent in Equation (\ref{eq:re_d_to_re_tau}) is found to be $12/13=0.9231$ in \cite{anbarlooei_a}, which deviates less than 1\% from our result.

The local power-law fit was done for $60 < z^+ < 0.15 Re_{\tau}$ and the semi-global power-law fit covered a range $40 < z^+ < 0.85 Re_{\tau}$. The local power-law fit can be compared to a range of $3\sqrt{Re_{\tau}} < z^+ < 15Re_{\tau}$ used for the local log-law fits in \cite{marusic_a}.

As mentioned in \cite{zagarola_a}, the local power-law exponent in Equation (\ref{eq:zag_l}), 0.137, is close to the 1/7 exponent ("the 1/7th law") proposed by von K\'arm\'an and Prandtl as a global power-law for the low Reynolds number mean velocity (see Chapter 2.4 in \cite{davidson_a} for the historical context).

It is interesting to note that the semi-global and global power-law parameters scale with $Re_{\tau}$ in a similar fashion; the difference is mainly an offset, which is due to the different radial ranges used for the fits.

\section{Discussion}
\label{sec:discussion}

\subsection{The high Reynolds number transition placed in context}

\begin{figure}[!ht]
\centering
\includegraphics[width=12cm]{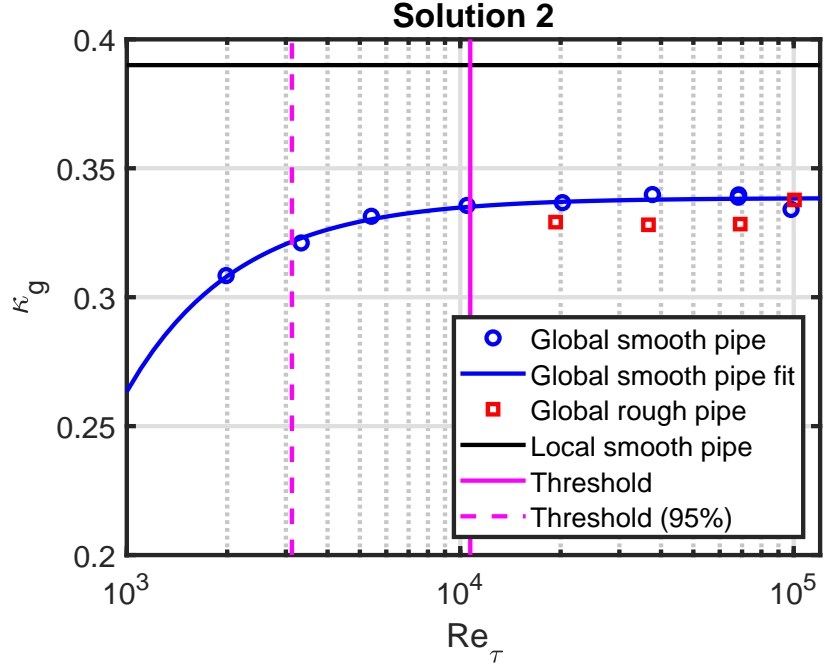}
\caption{$\kappa_{g,{\rm solution~2}}$ vs. $Re_{\tau}$. The transitional $Re_{\tau}$ using the 99\% (95\%) criterion is shown as the vertical solid (dashed) magenta line, respectively.}
\label{fig:sol2_kappa_95perc}
\end{figure}

As stated, we have identified a transition in mean flow behaviour around $Re_{\tau} \rvert_{\rm threshold}=10715$ at the point where $\kappa_{g,{\rm solution~2}}$ reached 99\% of the asymptotic value. This criterion is of course arbitrary to some extent; if we instead require 95\%, the transitional $Re_{\tau}$ is 3126, a factor of three lower, see Figure \ref{fig:sol2_kappa_95perc}. Therefore a single transitional Reynolds number is difficult to pin down, indicating that there is a gradual rather than an abrupt transition.

The transition we have found in this paper is consistent with earlier indications of transition, see Appendix D in \cite{russo_a}.

In \cite{marusic_c}, it is shown that the turbulent kinetic energy production in the logarithmic region exceeds the near-wall production above $Re_{\tau} \approx 4200$. This could be linked to the transition we are describing in this paper.

Global averaging has previously been used in \cite{yakhot_a}, where the streamwise component of the area-averaged mean turbulent kinetic energy is calculated:

\begin{equation}
  K = \bigg \langle \frac{\overline{u^2}}{2} \bigg \rangle_{\rm AA},
\end{equation}

\noindent which can be related to the area-average of the square of the normalised fluctuating velocity used in this paper:

\begin{equation}
  \bigg \langle \frac{\overline{u_g^2}}{U_{\tau}^2} \bigg \rangle_{\rm AA} = \frac{2K}{U_{\tau}^2}
\end{equation}

As a consistency check, we can compare results for $K/U_{\tau}^2$ in \cite{yakhot_a} with values we have calculated: They find that the ratio is in the range 1.6-1.7 above $Re_D \approx 10^5$. Multiplying this by two to convert to $\langle \overline{u_g^2}/U_{\tau}^2 \rangle_{\rm AA}$, we find a range 3.2-3.4 which agrees fairly well with the average value of the Superpipe measurements, namely 3.2, see Figure \ref{fig:turb_sq_vs_Re_tau} and Equations (\ref{eq:fluc_log}) and (\ref{eq:fluc_pow}).

\begin{figure}[!ht]
\centering
\includegraphics[width=12cm]{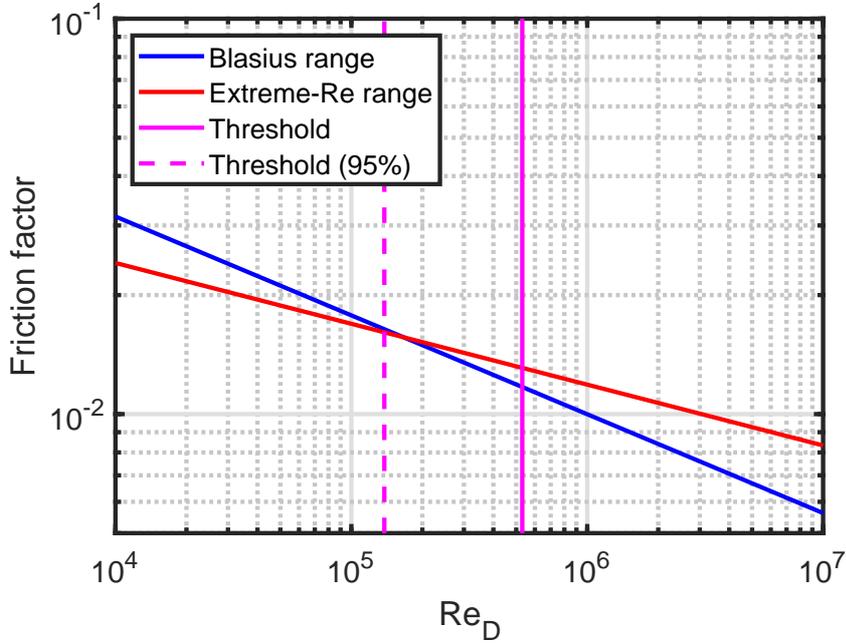}
\caption{Friction factor as a function of $Re_D$ for smooth pipes. The transitional $Re_D$ using the 99\% (95\%) criterion is shown as the vertical solid (dashed) magenta line, respectively.}
\label{fig:ff_scaling}
\end{figure}

A transition in the friction factor from Blasius scaling to "Extreme-Re" scaling has been identified for smooth pipes in \cite{anbarlooei_a}:

\begin{align}
  \lambda_{\rm Blasius} & = \frac{3.16 \times 10^{-1}}{Re_D^{1/4}} \label{eq:ff_Blasius} \\
  \lambda_{\rm Extreme-Re} & = \frac{9.946 \times 10^{-2}}{Re_D^{2/13}} \label{eq:ff_extreme},
\end{align}

\noindent see Figure \ref{fig:ff_scaling}. The nominal value for the transition is $Re_D = 166418$. A universal model for the friction factor in smooth pipes has been published in \cite{dixit_a}.

Under certain assumptions, the Blasius friction factor scaling can be used to derive the global 1/7th law, see Division G, Section 22 in \cite{durand_a}:

\begin{equation}
\frac{U_{\rm g{\rm -Blasius}}^2(z)}{U_{\tau}^2} = \left[ 8.74 \times (z^+)^{1/7} \right]^2
\end{equation}

Results from \cite{marusic_c,yakhot_a,anbarlooei_a,skouloudis_a} and this paper are summarized in Table \ref{tab:trans_re_comp}. For all cases, both $Re_{\tau}$ and $Re_D$ are shown. The results agree within a factor of two except for \cite{skouloudis_a} and the "Threshold (99\%)" values which are roughly a factor of three higher. It seems likely that all results detect the same transition; however, from our analysis it appears that the transitional Reynolds number is higher than previously estimated, but consistent with \cite{skouloudis_a}.

\begin{table}[!ht]
\caption{Transition Reynolds number.} 
\centering 
\begin{tabular}{ccc} 
\hline\hline 
Source & $Re_{\tau}$ & $Re_D$ \\  
\hline 
Marusic et al. \cite{marusic_c} & 4200 & 1.9 $\times 10^5$ \\
Yakhot et al. \cite{yakhot_a} & 2329 & 1.0 $\times 10^5$ \\
Anbarlooei et al. \cite{anbarlooei_a} & 3711 & 1.7 $\times 10^5$ \\
Skouloudis et al. \cite{skouloudis_a} & $10^4$ & 4.9 $\times 10^5$ \\
Threshold (95\%) & 3126 & 1.4 $\times 10^5$ \\
Threshold (99\%) & 10715 & 5.3 $\times 10^5$ \\
\hline 
\end{tabular}
\label{tab:trans_re_comp} 
\end{table}

\subsection{On the properties of the von K\'arm\'an constant}

\begin{figure}[!ht]
\centering
\includegraphics[width=6.5cm]{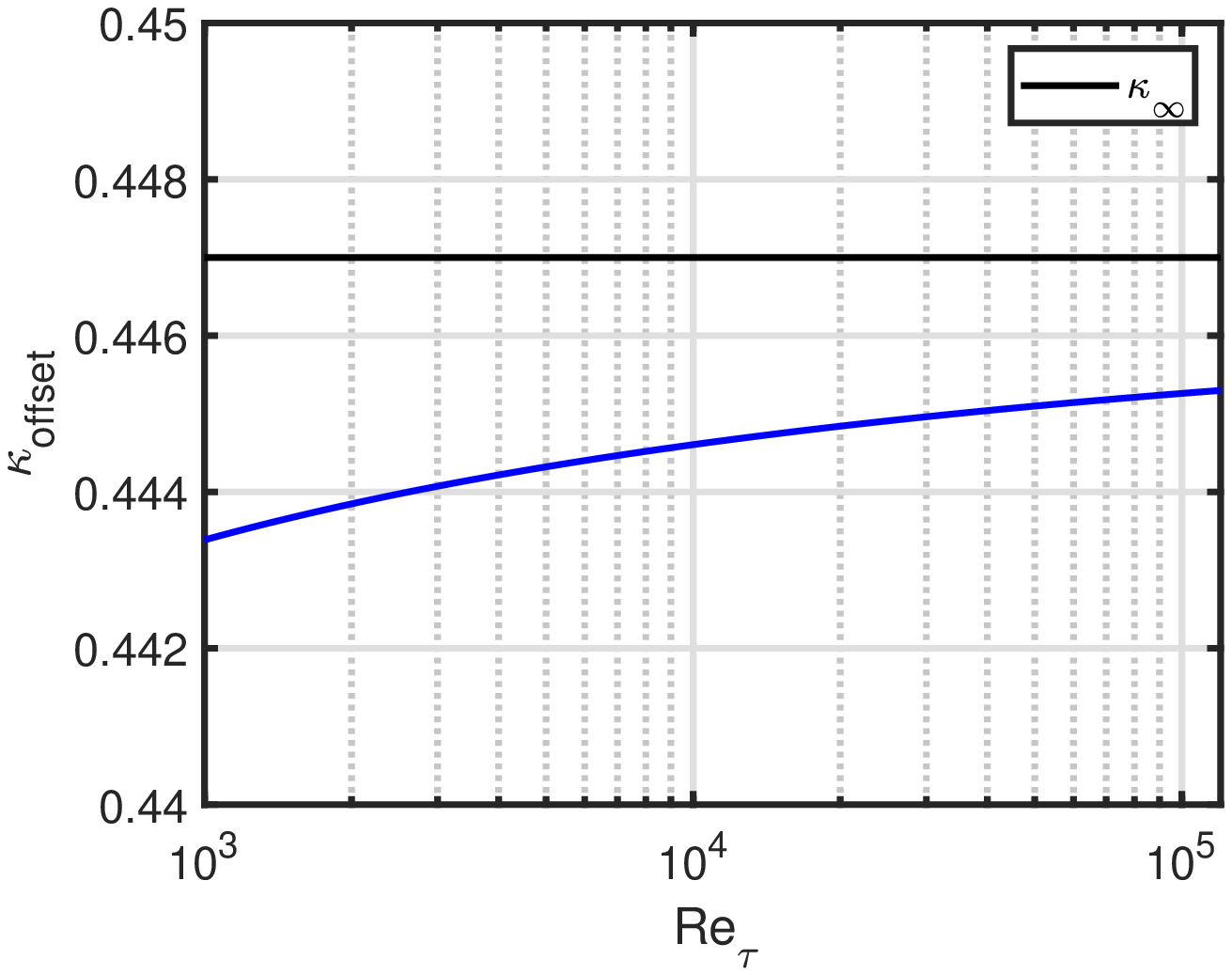}
\includegraphics[width=6.5cm]{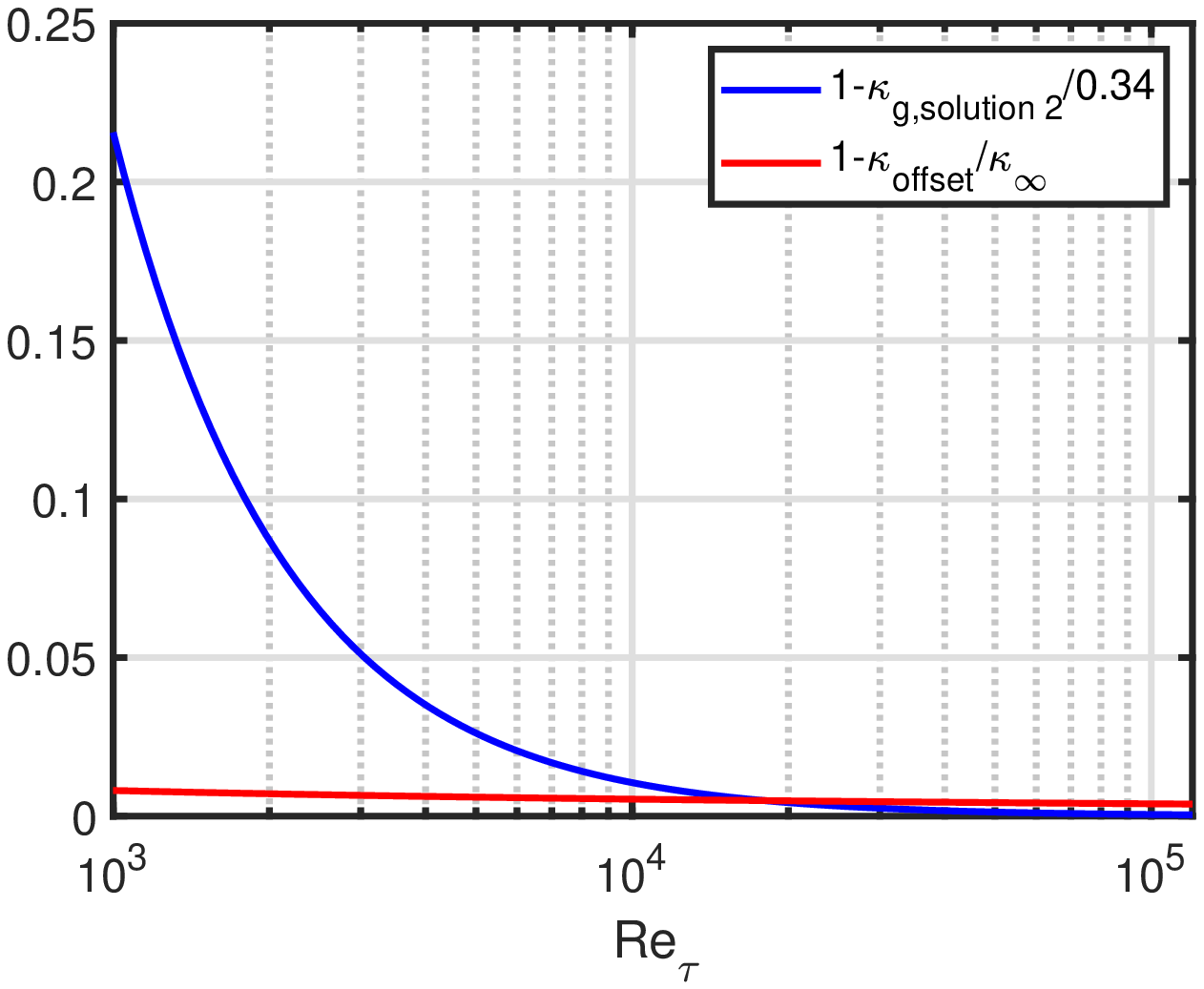}
\caption{Left-hand plot: $\kappa_{\rm offset}$ vs. $Re_{\tau}$, right-hand plot: Comparison between normalised change of $\kappa_{g,{\rm solution~2}}$ and $\kappa_{\rm offset}$.}
\label{fig:kappa_offset}
\end{figure}

A theory has been proposed where $z+z_{\rm offset}$ is considered instead of $z$, $z_{\rm offset}$ being an offset associated with a mesolayer \cite{wosnik_a}. One consequence is the variation of the von K\'arm\'an constant with Reynolds number:

\begin{equation}
\kappa_{\rm offset} = \frac{\kappa_{\infty} (\log(D_s Re_{\tau}))^{1+\alpha}}{(\log(D_s Re_{\tau}))^{1+\alpha}-\alpha A \kappa_{\infty}},
\end{equation}

\noindent which is shown in the left-hand plot of Figure \ref{fig:kappa_offset} for $\kappa_{\infty}=0.447$, $D_s=1$, $\alpha=0.44$ and $A=-0.67$. The trend is similar to what we have observed, but the approach to the asymptotic value of the von K\'arm\'an constant is slower for the offset theory, see the right-hand plot of Figure \ref{fig:kappa_offset}.

Variations of the von K\'arm\'an constant with Reynolds number have been compared for different canonical flows in \cite{nagib_a}. The von K\'arm\'an constant approaches an asymptotic value, but the value of this number is different for the three flow types. These results and others are included in \cite{marusic_d} and it is stated that the von K\'arm\'an constant approaches the asymptotic value from below for pipe flow, which matches what we have found.

The concepts of active and inactive vortex motion \cite{townsend_a,deshpande_a} relates to effects of smaller near-wall and larger core eddies. In \cite{davidson_b}, a possible consequence of this distinction is shown to be scaling of the von K\'arm\'an constant with Reynolds number. The outcome is that the measured value of the von K\'arm\'an constant is larger than the universal value by a factor which depends on the streamwise TI squared among other terms.

\subsection{Length scales}
\label{subsec:length_scales}

A characteristic mixing length scale $l_m$ (see Chapter 2.5 in \cite{davidson_a}) can be found by differentiating the mean velocity log-law:

\begin{equation}
\frac{d U^+}{dz} = \frac{1}{\kappa z} = \frac{1}{l_m},
\end{equation}

\noindent which shows that the scale of turbulence is proportional to the distance from the wall (attached eddies \cite{townsend_a}):

\begin{equation}
l_m = \kappa z
\end{equation}

Since we have found that $\kappa_g < \kappa_l$, the global characteristic length scale $l_g=\kappa_g z$ is smaller than the corresponding local length scale $l_l=\kappa_l z$ for a given distance from the wall. We also note that if the von K\'arm\'an constant increases with Reynolds number, the turbulent structure size increases correspondingly.

Integral length scales can be found by calculating the AM and AA of the mixing length:

\begin{equation}
l_{i,~{\rm AM}}=\langle l_m \rangle_{\rm AM} = \frac{1}{\delta} \times \int_{0}^{\delta} \kappa z {\rm d}z = \frac{\kappa \delta}{2}
\end{equation}

\begin{equation}
l_{i,~{\rm AA}}=\langle l_m \rangle_{\rm AA} = \frac{2}{\delta^2} \times \int_{0}^{\delta} \kappa z \times (\delta-z) {\rm d}z = \frac{\kappa \delta}{3}
\end{equation}

In \cite{schlichting_a}, a turbulent length scale is derived from the normalised correlation function $R(z)$ to be:

\begin{equation}
l_c = \int_0^{\delta} R(z) {\rm d}z \approx 0.14 \delta
\end{equation}

If we require $l_{i,~{\rm AM}}=l_c$ ($l_{i,~{\rm AA}}=l_c$), we get $\kappa=0.28$ $(0.42)$, respectively, which is comparable to the range of values between $\kappa_g$ and $\kappa_l$.

Our discussion pertains to wall-normal length scales; spanwise and/or streamwise structures can extend to lengths far greater than the pipe diameter.

\subsection{Radial profiles of turbulence intensity}
\label{sec:rad_prof}

The global log-law TI profile is defined as in Equation (\ref{eq:TI_sq_def}), but with the subscript changed from "l" to "g". This is shown for solution 2 as "Global log-law, solution 2" in Figure \ref{fig:I_sq_vs_alpha_lo_hi_Re}. In this figure the "Global power-law" is also present, defined as:

\begin{equation}
I^2_g(z) \rvert_{\rm power-law} = \frac{a_g \times \left( \frac{z}{\delta} \right)^{b_g}}{c_g \times \left( z^+ \right)^{d_g}}
\end{equation}

\subsection{Turbulence intensity scaling with Reynolds number}

Below we define the TI squared using ratios for either log- or power-laws separately.

\begin{figure}[!ht]
\centering
\includegraphics[width=12cm]{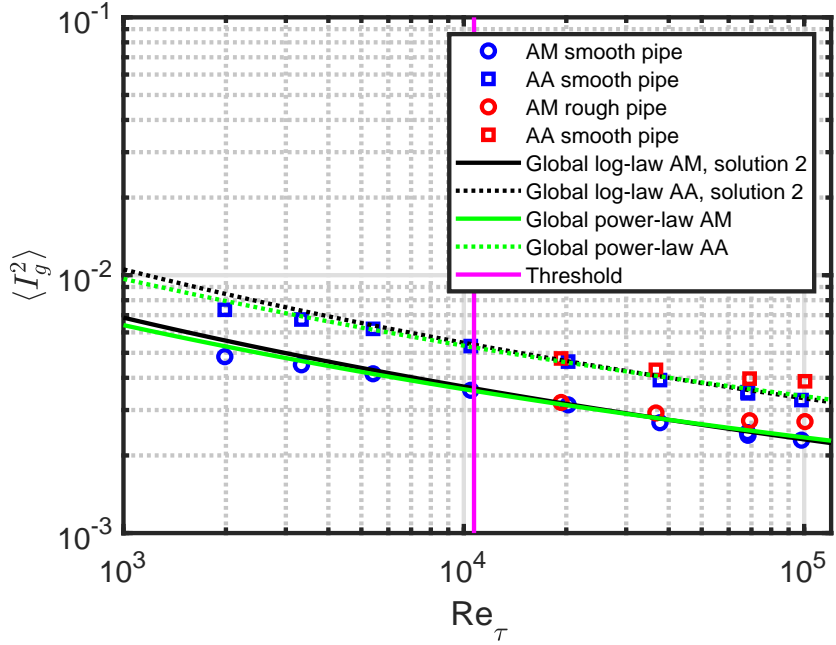}
\caption{Averaged global TI squared as a function of $Re_{\tau}$.}
\label{fig:I_sq_vs_Re}
\end{figure}

The averaged global TI squared is defined below for AM and AA, see Figure \ref{fig:I_sq_vs_Re}:

\begin{equation}
\langle I^2_g \rangle_{\rm AM} = \biggl \langle \frac{{\overline{u^2_g}}}{U_{\tau}^2} \biggr \rangle_{\rm AM} \bigg/ \biggl \langle \frac{U^2_g}{U_{\tau}^2} \biggr \rangle_{\rm AM}
\end{equation}

\begin{equation}
\langle I^2_g \rangle_{\rm AA} = \biggl \langle \frac{{\overline{u^2_g}}}{U_{\tau}^2} \biggr \rangle_{\rm AA} \bigg/ \biggl \langle \frac{U^2_g}{U_{\tau}^2} \biggr \rangle_{\rm AA}
\end{equation}

\subsubsection{Log-law}

The averaged global TI squared for the log-law is defined for AM using Equations (\ref{eq:avrg_fluc_log_AM}) and (\ref{eq:avrg_mean_log_AM}):

\begin{equation}
\langle I^2_g \rangle_{\rm AM,log-law} = \frac{B_{1,g}+A_{1,g}}{\frac{2}{\kappa^2_g} - \frac{2A_g}{\kappa_g} + A^2_g + \log(Re_{\tau}) \left( \frac{2A_g}{\kappa_g} - \frac{2}{\kappa^2_g} \right) + \frac{\log^2(Re_{\tau})}{\kappa^2_g}}
\end{equation}

The averaged global TI squared for the log-law is defined for AA using Equations (\ref{eq:avrg_fluc_log_AA}) and (\ref{eq:avrg_mean_log_AA}):

\begin{equation}
\langle I^2_g \rangle_{\rm AA,log-law} = \frac{B_{1,g} + \frac{3}{2} \times A_{1,g}}{\frac{7}{2\kappa^2_g} - \frac{3A_g}{\kappa_g} + A^2_g + \log(Re_{\tau}) \left( \frac{2A_g}{\kappa_g} - \frac{3}{\kappa^2_g} \right) + \frac{\log^2(Re_{\tau})}{\kappa^2_g}}
\end{equation}

These definitions are shown as black lines in Figure \ref{fig:I_sq_vs_Re}. Mean values of $A_{1,g}$ and $B_{1,g}$ from Equations (\ref{eq:cst_A_1g}) and (\ref{eq:cst_B_1g}) have been used.

\subsubsection{Power-law}

The averaged global TI squared for the power-law is defined for AM using Equations (\ref{eq:avrg_fluc_pow_AM}) and (\ref{eq:avrg_mean_pow_AM}):

\begin{equation}
\langle I^2_g \rangle_{\rm AM,power-law} = \frac{a_g}{b_g+1} \times \frac{d_g+1}{c_g \times Re_{\tau}^{d_g}}
\end{equation}

The averaged global TI squared for the power-law is defined for AA using Equations (\ref{eq:avrg_fluc_pow_AA}) and (\ref{eq:avrg_mean_pow_AA}):

\begin{equation}
\langle I^2_g \rangle_{\rm AA,power-law} = \frac{2a_g}{(b_g+1)(b_g+2)} \times \frac{(d_g+1)(d_g+2)}{2c_g \times Re_{\tau}^{d_g}}
\end{equation}

These definitions are shown as green lines in Figure \ref{fig:I_sq_vs_Re}.

\subsection{Turbulence intensity scaling with friction factor}

\begin{figure}[!ht]
\centering
\includegraphics[width=12cm]{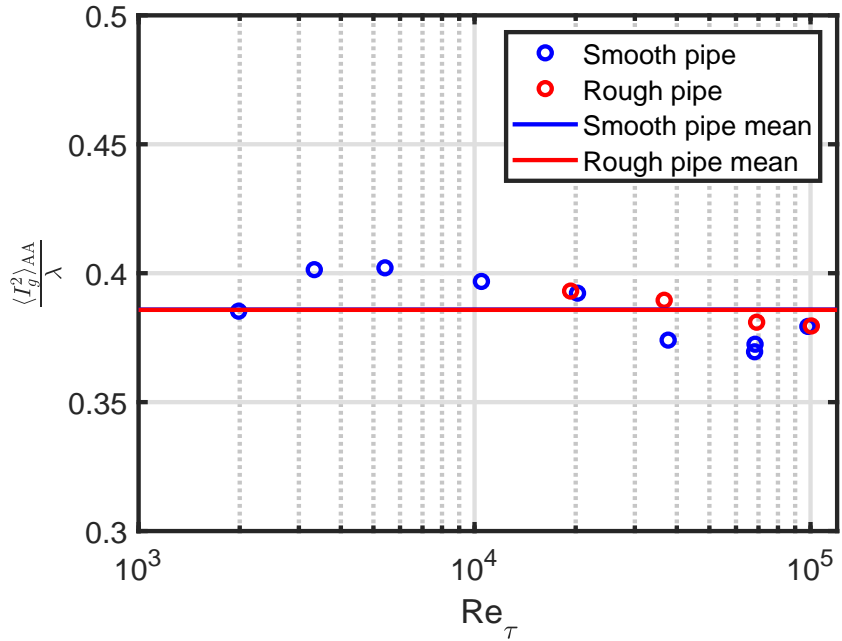}
\caption{$\langle I^2_g \rangle_{\rm AA}/\lambda$ as a function of $Re_{\tau}$. The smooth- and rough-wall results are in the same range.}
\label{fig:I_sq_AA_div_lambda_vs_Re_tau}
\end{figure}

We have presented results relating the TI and the friction factor in \cite{basse_b,basse_c} and we can discuss it further based on our findings in this paper. The friction factor is defined as:

\begin{equation}
\lambda=8 \frac{U_{\tau}^2}{\langle U^2_g \rangle_{\rm AA}} = \frac{8}{\biggl \langle \frac{U^2_g}{U_{\tau}^2} \biggr \rangle_{\rm AA}},
\end{equation}

\noindent which can be rewritten as:

\begin{align}
  \langle I^2_g \rangle_{\rm AA} &= \frac{\lambda}{8} \times \biggl \langle \frac{{\overline{u^2_g}}}{U_{\tau}^2} \biggr \rangle_{\rm AA} \\
   &= \frac{\lambda}{8} \times \left( B_{1,g} + \frac{3}{2} \times A_{1,g} \right) \\
   &= \frac{\lambda}{8} \times \frac{2a_g}{(b_g+1)(b_g+2)} \\
   &= 0.39 \times \lambda \\
   &= \kappa_{l} \times \lambda \label{eq:TI_lambda},
\end{align}

\noindent leading to a simple relationship between the area-averaged, squared TI and the friction factor in the final Equation (\ref{eq:TI_lambda}): They are proportional with the local von K\'arm\'an constant $\kappa_l$ as the factor of proportionality. This relationship is illustrated in Figure \ref{fig:I_sq_AA_div_lambda_vs_Re_tau}. It is generally valid, e.g. for all $Re_{\tau}$ and smooth- and rough-wall pipe flow.

We end by combining Equations (\ref{eq:ff_Blasius}) and (\ref{eq:ff_extreme}) with Equation (\ref{eq:TI_lambda}) to derive expressions for the area-averaged, squared TI explicitly as a function of $Re_D$:

\begin{align}
  \langle I^2_g \rangle_{\rm AA,~Blasius} &= \kappa_{l} \times \frac{3.16 \times 10^{-1}}{Re_D^{1/4}} \label{eq:smooth_pred_Blasius} \\
  \langle I^2_g \rangle_{\rm AA,~Extreme-Re} &= \kappa_{l} \times \frac{9.946 \times 10^{-2}}{Re_D^{2/13}} \label{eq:smooth_pred_extreme}
\end{align}

\begin{figure}[!ht]
\centering
\includegraphics[width=12cm]{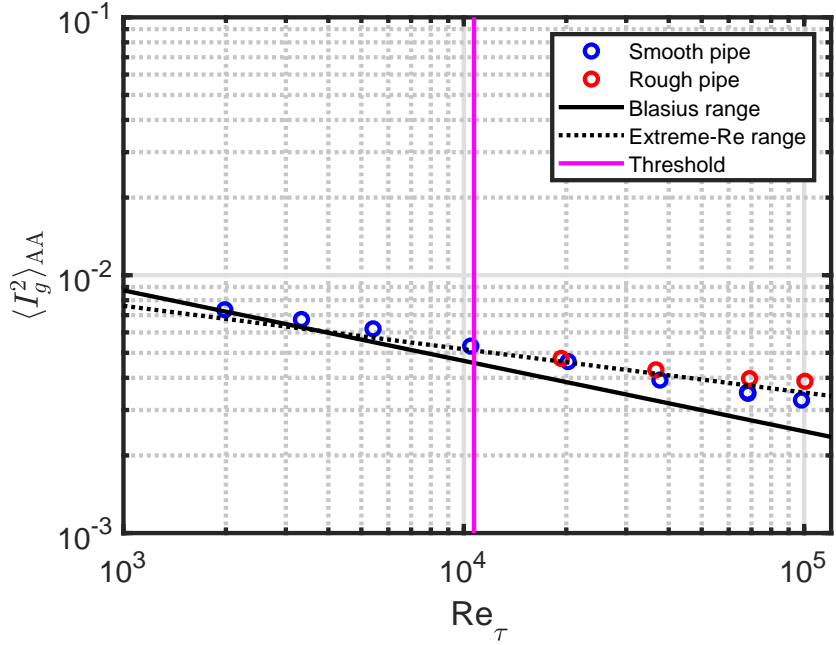}
\caption{$\langle I^2_g \rangle_{\rm AA}$ as a function of $Re_{\tau}$. The solid (Blasius) and dotted (Extreme-Re) lines are the smooth pipe predictions from Equations (\ref{eq:smooth_pred_Blasius}) and (\ref{eq:smooth_pred_extreme}).}
\label{fig:I_sq_AA_smooth_pred}
\end{figure}

The performance of these expressions is shown in Figure \ref{fig:I_sq_AA_smooth_pred}. Note that the friction factor scaling used \cite{anbarlooei_a} is for smooth pipes. The agreement with measurements is reasonable, although the slope for the "Extreme-Re" range is not matching the smooth pipe measurements exactly.

\subsection{Alternative solutions to the global log-law}

We have focused on solution 2 for the global mean velocity log-law. Here, we return briefly to the other solutions, see Figure \ref{fig:solutions_lo_hi_Re}.

Solution 1 is simply flow in the reverse direction of solution 2.

Solutions 3 and 4 have zero crossings at much higher $z^+$ than the local log-law. If one would interpret this in a physical sense, it could be equivalent to Couette flow with walls moving in opposite directions, see e.g. \cite{tillmark_a}.

\begin{figure}[!ht]
\centering
\includegraphics[width=6.5cm]{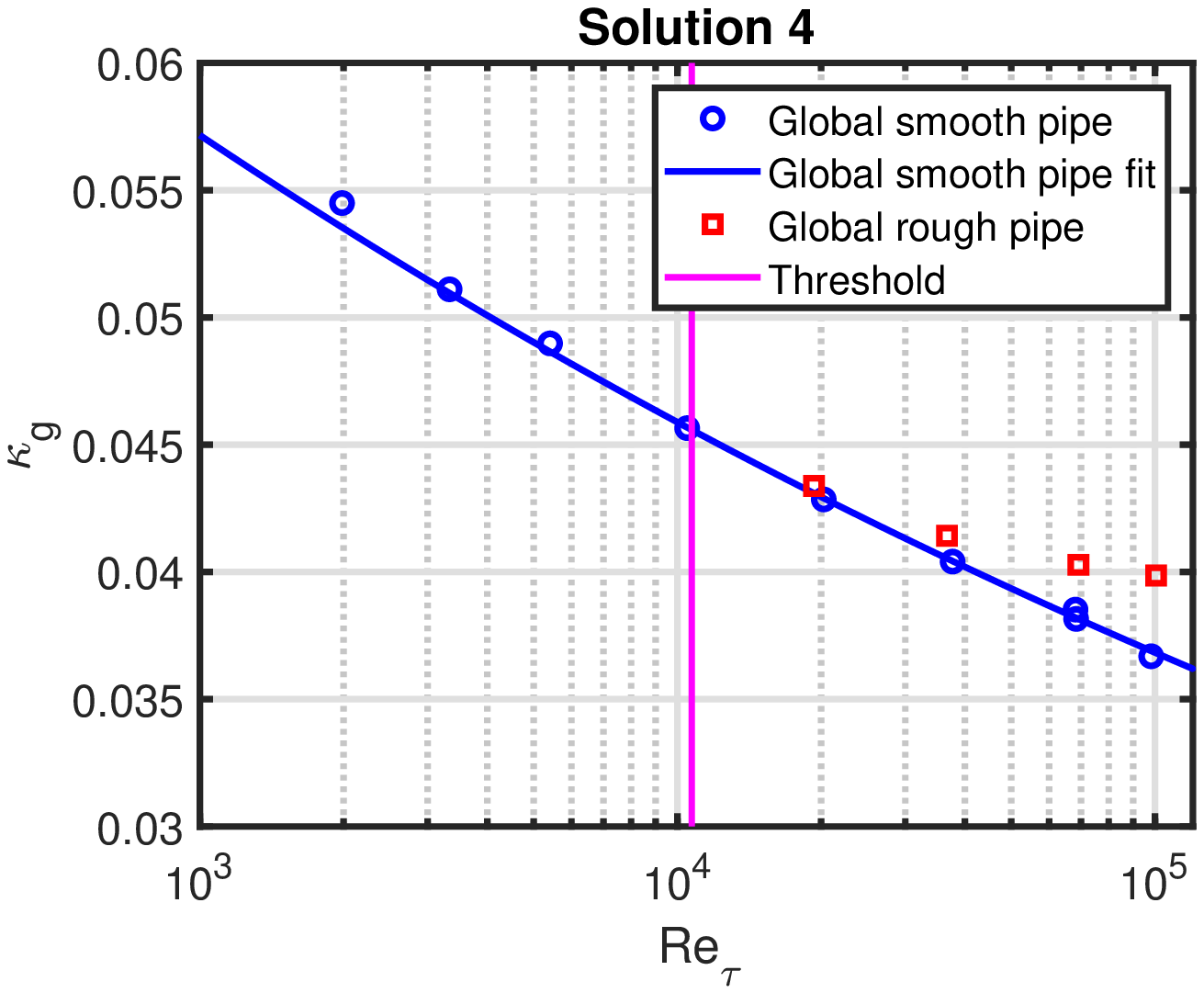}
\includegraphics[width=6.5cm]{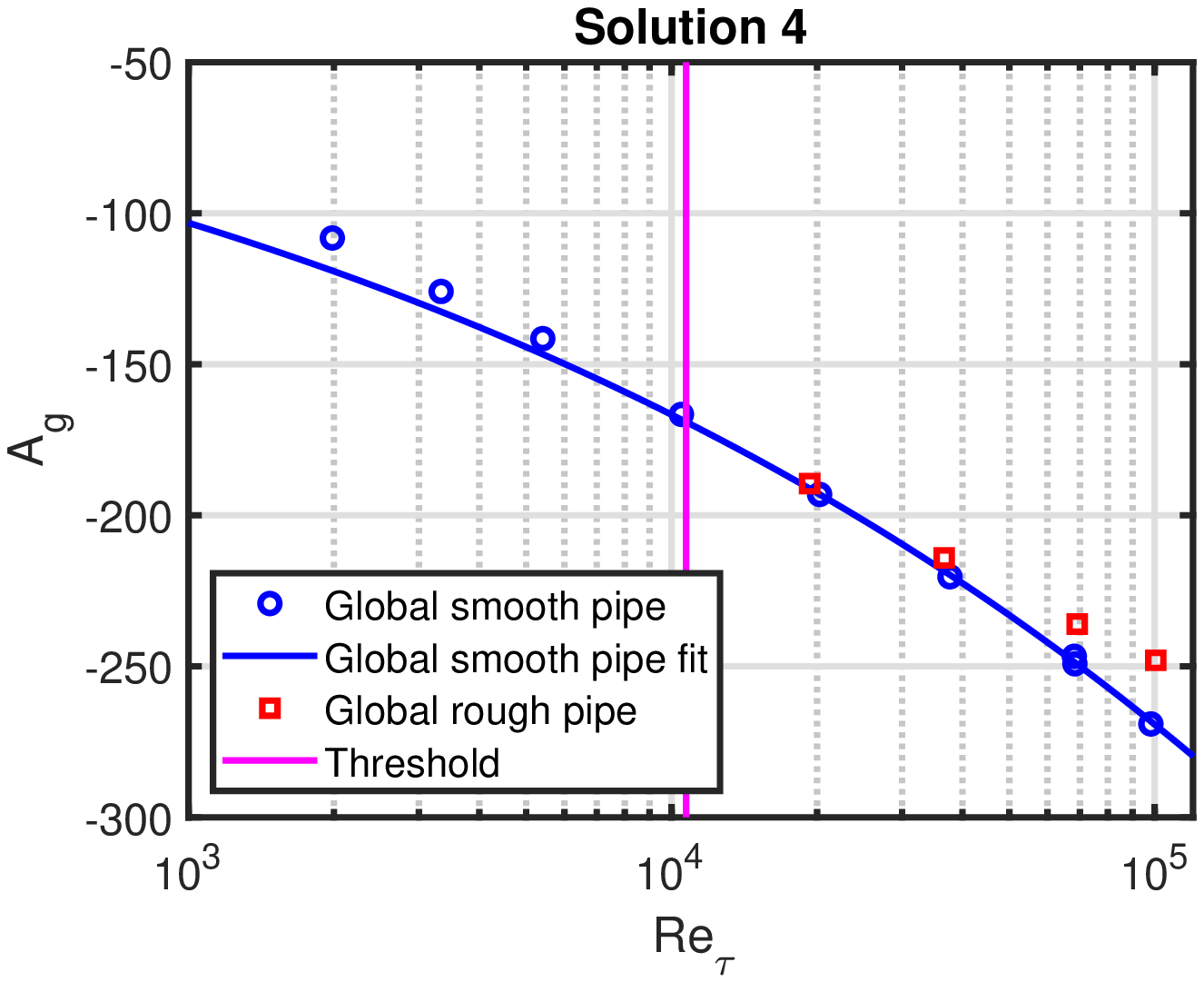}
\caption{Left-hand plot: $\kappa_{g,{\rm solution~4}}$ vs. $Re_{\tau}$, right-hand plot: $A_{g,{\rm solution~4}}$ vs. $Re_{\tau}$. Blue lines are Equations (\ref{eq:sol4_kappa}) and (\ref{eq:sol4_A}).}
\label{fig:solution4_kappa_A}
\end{figure}

For completeness, we characterise solution 4. Both $A_{g,{\rm solution~4}}$ and $\kappa_{g,{\rm solution~4}}$ are functions of Reynolds number. Fits for Reynolds numbers above the threshold yield:

\begin{eqnarray}
  A_{g,{\rm solution~4}} &=& -24.49 \times Re_{\tau}^{0.21} \qquad R^2=0.99 \label{eq:sol4_kappa} \\
  \kappa_{g,{\rm solution~4}} &=& 0.11 \times Re_{\tau}^{-0.10} \qquad R^2=0.99 \label{eq:sol4_A},
\end{eqnarray}

\noindent see Figure \ref{fig:solution4_kappa_A}. Thus, it adds some complexity compared to solution 2 where both parameters are constant above the threshold Reynolds number. However, it is a mathematically valid solution as well.

\subsection{Higher order radial averaging}

We have considered equations for fluctuating and mean velocities defined using two parameters. Thus, two equations for radial averaging were required, AM and AA.

If we were to consider e.g. equations using three parameters, we would need a third radial averaging equation, volume averaging (VA) \cite{basse_c}.

For this case the square of the fluctuating and mean velocities could be expressed as:

\begin{equation}
\frac{{\overline{u^2_g}}(z)}{U_{\tau}^2} = \alpha_g \times \left( \frac{z}{\delta} \right)^{\beta_g} + \gamma_g
\end{equation}

\noindent and:

\begin{equation}
\frac{U_g^2(z)}{U_{\tau}^2} = \varepsilon_g \times \left( z^+ \right)^{\zeta_g} + \eta_g,
\end{equation}

\noindent respectively, where $\alpha_g, \beta_g, \gamma_g, \varepsilon_g, \zeta_g$ and $\eta_g$ are parameters.

Other power-law and log-law solutions, inspired by the offset solution \cite{wosnik_a} and grid-generated turbulence decay (see e.g. Chapter 3.3.1 in \cite{davidson_a}), could be:

\begin{eqnarray}
  \frac{{\overline{u^2_g}}(z)}{U_{\tau}^2} &=& \alpha_g \times \left( \frac{z}{\delta} + \gamma_g \right)^{\beta_g} \\
  \frac{U_g^2(z)}{U_{\tau}^2} &=& \varepsilon_g \times \left( z^+ + \eta_g \right)^{\zeta_g} \\
  \frac{{\overline{u^2_g}}(z)}{U_{\tau}^2} &=& \alpha_g - \beta_g \times \log \left( \frac{z}{\delta} + \gamma_g \right) \\
  \frac{U_g^2(z)}{U_{\tau}^2} &=& ( \zeta_g \times \log (z^+ + \eta_g) + \varepsilon_g )^2
\end{eqnarray}

This is outside the scope of the current paper but will be addressed in future research.

\subsection{Scaling of the peak of the squared normalised fluctuating velocity}

We note that a radial redistribution of the velocity fluctuations as a function of $Re_{\tau}$ might occur - but not be detected - due to the averaging process.

\begin{figure}[!ht]
\centering
\includegraphics[width=12cm]{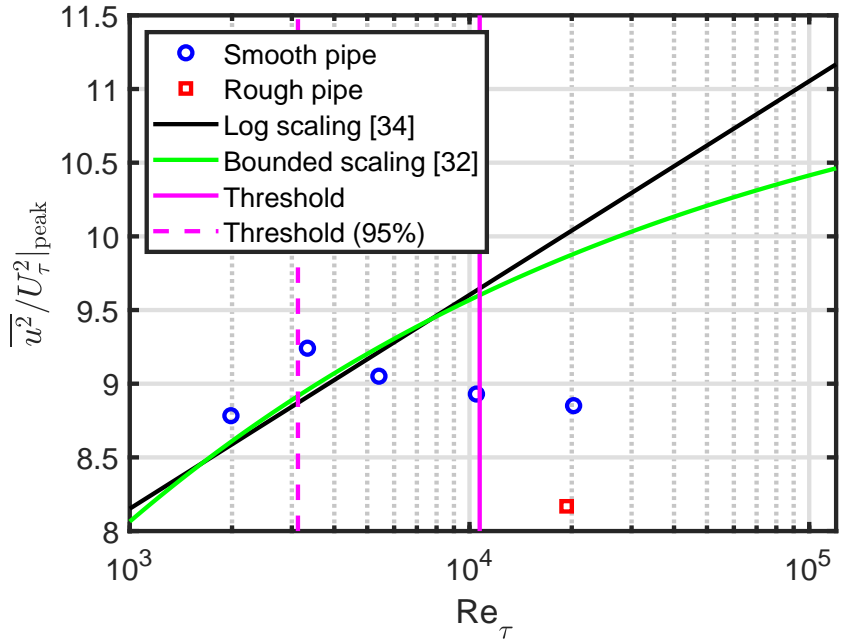}
\caption{$\overline{u^2}/U_{\tau}^2 \rvert_{\rm peak}$ as a function of $Re_{\tau}$. The solid black line is from \cite{marusic_e} and the solid green line from \cite{chen_a}. The transitional $Re_{\tau}$ using the 99\% (95\%) criterion is shown as the vertical solid (dashed) magenta line, respectively. Superpipe measurements are included for the Reynolds numbers where a distinct inner peak is visible.}
\label{fig:peak_scaling}
\end{figure}

We know that the peak of the squared normalised fluctuating velocity scales with $Re_{\tau}$, but recent work \cite{chen_a,monkewitz_a} indicates that the peak becomes asymptotically constant (bounded). This is in contrast to previous scaling expressions where the peak is proportional to the logarithm of the Reynolds number \cite{marusic_e}, see Figure \ref{fig:peak_scaling}. Using the expression for $\overline{u^2}/U_{\tau}^2 \rvert_{\rm peak}$ in \cite{chen_a}, the peak value is 9.6 for $Re_{\tau}=10715$, which is 20\% below the asymptotic value of 11.5. The Princeton Superpipe data we treat is part of the measurement database used in \cite{chen_a}; it is interesting to note that the Superpipe peak values appear to become asymptotic at a value of roughly 9 for $Re_{\tau}$ above $4000$. This value matches the transitional Reynolds numbers in Table \ref{tab:trans_re_comp} quite well.

\subsection{Recommendations}

As stated in \cite{pope_a}, the log-law has the advantage over the power-law of being universal, i.e. independent of Reynolds number for asymptotically high Reynolds numbers.

The question of log-law versus power-law behaviour has been debated for more than a century, with bursts of publications followed by periods of relative calm. One such burst occurred in relation to the publication of \cite{barenblatt_a,barenblatt_b}, where a power-law approach was advocated, followed by various rebuttals, e.g. \cite{zagarola_a}.

We have seen that log-laws and power-laws characterise the Superpipe measurements equally well, but the universality of the log-laws puts them at a slight advantage. To quote \cite{davidson_b}, "Occam's razor might lead us to favour the log-law.".

\section{Conclusions}
\label{sec:conclusions}

By an analysis of global properties of fluctuating and mean pipe flow velocities, we have characterised a high Reynolds number transition region at a friction Reynolds number $Re_{\tau} \sim 11000$. The transitional Reynolds number appears slightly higher than reported in literature, so the global von K\'arm\'an constant $\kappa_g$ may be a more sensitive indicator than those used previously. A consequence of this transition is that we cannot use a single scaling expression for turbulent flow across the entire Reynolds number range. This is important for CFD and other industrial applications.

Fluctuating and mean velocities have been treated separately and combined to calculate the TI. Scaling with Reynolds number and the impact of wall roughness is only seen for the mean flow. We have applied a novel method to derive two-parameter radial expressions using both log-law and power-law functions; they capture the main features of the Princeton Superpipe measurements equally well. However, we would tend to recommend using log-law functions since they become universal for high Reynolds numbers.

We have shown that the area-averaged square of the TI is proportional to the friction factor, the proportionality constant being the local von K\'arm\'an constant $\kappa_l=0.39$.

\paragraph{Acknowledgements}

We thank Professor Alexander J. Smits for making the Princeton Superpipe data publicly available.

\paragraph{Data availability statement}

Data sharing is not applicable to this article as no new data were created or analyzed in this study.


\label{sec:refs}

\end{document}